\documentclass{jfm}

\usepackage{graphicx,amsmath,amssymb}

\usepackage{epstopdf, epsfig}
\usepackage{natbib}
\usepackage{anyfontsize}

\newcommand{\infd}{\text{d}}
\newcommand{\diff}[2]{\frac{\infd{#1}}{\infd{#2}}}
\newcommand{\pdiff}[2]{\frac{\partial{#1}}{\partial{#2}}}

\newcommand{\V}{\mathcal{V}}

\title{Diapycnal material transport driven by submesoscale frontogenesis}

\author{Tong Bo\aff{1,2}
  \corresp{\email{tong.bo@pku.edu.cn}},
  James C. McWilliams\aff{2}
 \and Marcelo Chamecki\aff{2}}

\affiliation{\aff{1} School of Mechanics and Engineering Science, 
Peking University, Beijing 100871, 
People’s Republic of China
\aff{2} Department of Atmospheric and Oceanic Sciences, 
University of California Los Angeles, Los Angeles,
CA 90095-1565, USA}

\begin{document}

\maketitle

\begin{abstract}
     Submesoscale fronts, occurring at intermediate scales between mesoscale 
     eddies and boundary layer turbulence, play a crucial role in driving 
     vertical transport from the ocean surface into the interior. 
     Their dynamics involve complex interactions between 
     submesoscale currents and turbulence. However, the mechanisms by 
     which these multiscale processes combine to transport tracers such 
     as pollutants or nutrients remain less well understood. 
     This study uses large-eddy simulation to investigate 
     passive tracer transport associated with submesoscale fronts. 
     Intense turbulence develops during frontogenesis, leading to strong 
     diapycnal tracer transport into the ocean interior. 
     While part of this transport arises from the direct turbulent flux, 
     represented by the covariance between turbulent velocity and tracer 
     concentration fluctuations, a substantial portion is due to an 
     advective diapycnal flux driven by the mean diapycnal velocity. 
     The mean diapycnal velocity results from the evolving secondary 
     circulation in the presence of turbulent density mixing. 
     These findings reveal an underexplored diapycnal transport pathway 
     in submesoscale frontal zones, with implications for improved 
     representation of vertical exchange in ocean models. 
\end{abstract}

\section{Introduction}
In the ocean, the global thermohaline circulation relies on 
diapycnal fluxes of buoyancy and materials to complete its circuit. 
Diapycnal fluxes are typically weak within the stably stratified 
ocean interior due to inhibited turbulent mixing. 
Vertical fluxes are also limited in the surface mixed layer, 
where density gradients are small. As a result, attention is 
directed toward vertical transport at the interface between 
the mixed layer and the ocean interior. 
Submesoscale processes, which are usually associated with 
horizontally divergent surface currents and strong vertical 
velocities, play a critical role in mediating such vertical 
transport \citep{mahadevan2006,freilich2021,taylor2023,pham2024}. 

Submesoscale currents typically consist of coherent structures 
including fronts, filaments, and eddies, characterized 
by spatial scales on the order of $1$~km and timescales on 
the order of hours to days \citep[e.g.,][]{mcwilliams2016}. 
Submesoscale frontal zones are usually defined by strong 
density gradients in one horizontal direction (cross-front 
direction), while gradients in the perpendicular direction 
(along-front direction) are comparatively weaker 
\citep{mcwilliams2021}. In addition to single-sided fronts, 
submesoscale currents may appear as dense filaments, 
which can be viewed as two-sided fronts, distinguished by 
sharp density gradients on both sides of a local maximum 
\citep{mcwilliams2009}. Submesoscale processes interact 
closely with larger mesoscale motions, from which they extract 
energy, and contribute to the forward energy cascade toward 
smaller-scale turbulence \citep{callies2015, sullivan2018}. 

The systematic formation and intensification of submesoscale fronts is 
known as frontogenesis, which can be driven by mesoscale strain flows 
or by turbulent thermal wind (TTW) \citep{hoskins1972,mcwilliams2021}. 
A positive feedback mechanism often emerges during frontogenesis, 
where ageostrophic secondary circulation in the cross-front plane 
accelerates frontal sharpening through interactions among density 
gradients, vorticity, and horizontal divergence 
\citep{sullivan2018,barkan2019}. 
As submesoscale currents evolve nonlinearly, various instabilities 
may develop \citep{gula2022}. Frontal instability generates turbulence 
and cross-frontal fluxes of density and momentum that can arrest 
further sharpening \citep{mcwilliams2016,gula2022,taylor2023}. 
In the absence of continued energy input from larger-scale flows, 
the arrested submesoscale fronts will eventually decay through the 
forward energy cascade \citep{sullivan2024}. 
Turbulence usually plays an important role throughout this life cycle, 
from frontogenesis to the arrest and eventual decay of submesoscale 
currents \citep{sullivan2018,bodner2023,dauhajre2025}. 

Tracer transport from the surface mixed layer into the ocean 
interior in submesoscale frontal zones is driven by multiscale 
flows \citep{freilich2019,pham2024}. 
The downward and upward velocities resulting from ageostrophic 
secondary circulation can lead to advective tracer fluxes. 
Downwelling is typically more intense and spatially concentrated 
than upwelling, making submesoscale fronts 
critical pathways for subduction \citep{taylor2018,freilich2021}. 
Submesoscale filaments (two-sided fronts) often exhibit stronger 
surface convergence and downwelling compared to single-sided 
fronts \citep{mcwilliams2009,bo2025front}. 
In addition to advective transport, turbulence arising from 
both frontal dynamics (e.g., frontogenesis and frontal instability) 
and background boundary layer forcing (e.g., surface cooling, 
wind stress, or surface waves) can induce vertical velocities on the 
order of 1~cm/s \citep{sullivan2018,freilich2021}. 
Turbulent motions not only interact with frontal dynamics, but also 
contribute directly to vertical tracer transport through small-scale 
eddy covariance, also known as turbulent fluxes \citep{pham2024}. 

Understanding how multiscale processes contribute to tracer 
transport in submesoscale frontal zones remains a challenge, despite 
its importance for ocean biogeochemical cycles and the water 
mass characteristics of the ocean interior \citep{mahadevan2020}. 
In field observations, the lack of clear separation between 
along-isopycnal advection and vertical displacement of isopycnals 
complicates the interpretation of tracer pathways. 
In numerical models, previous simulations that did not explicitly 
resolve turbulence suggested that enhanced downward velocities 
associated with submesoscale currents can trap tracers beneath 
the surface mixed layer \citep{omand2015,freilich2021}. 
However, recent studies show that turbulence interacts strongly 
with submesoscale frontal dynamics \citep{sullivan2018,bo2025front}. 
This underscores the need for modeling approaches that can  
better capture the influences of turbulence, such as large-eddy 
simulation (LES) \citep[e.g.,][]{hamlington2014,verma2022}. 
Improved understanding of tracer transport thus requires integrating 
knowledge of submesoscale frontal dynamics with investigation of 
turbulence-driven fluxes. Yet due to the high computational cost of 
resolving large domains at sufficient resolution, LES studies of 
tracer transport in frontal zones remain limited. 

In this study, we use LES to explore the mechanisms of passive 
tracer transport in submesoscale frontal zones. 
In particular, we focus on processes driving tracer fluxes 
into the stratified ocean interior, associated with both 
frontal dynamics and turbulence. 
Section~\ref{sec:methods} describes the LES model setup and 
theoretical framework. 
Section~\ref{sec:results} presents the frontal dynamics and 
analyzes diapycnal transport. 
Section~\ref{sec:discussion} discusses the generality of 
the findings and section~\ref{sec:conclusion} presents the conclusion.

\section{Methods}
\label{sec:methods}

\subsection{LES formulation}
We use LES to study submesoscale frontal dynamics and the 
associated passive tracer transport processes. 
LES is chosen as it can effectively capture the multiscale 
interactions between submesoscale currents and turbulence, 
as well as their combined effects on tracer fluxes. 
The LES framework is based on a set of grid‐filtered equations 
for mass, momentum, heat, and passive tracer 
\citep[e.g.,][]{deardorff1970,smagorinsky1963,bou2005,chamecki2008}: 
\begin{equation}
\label{eq:mass}
    \nabla\cdot\tilde{\boldsymbol u}=0;
\end{equation}
\begin{equation}
\label{eq:mom}
    \pdiff{\tilde{\boldsymbol u}}{t} 
+ \tilde{\boldsymbol u}\cdot\nabla\tilde{\boldsymbol u} 
 =  -\nabla\tilde{p}^* - f\boldsymbol{e}_z\times\tilde{\boldsymbol u} 
+ \tilde b\boldsymbol{e}_z 
 - \nabla\cdot\boldsymbol{\tau}^{sgs};
\end{equation}
\begin{equation}
\label{eq:heat}
    \pdiff{\tilde{\theta}}{t}+\tilde{\boldsymbol u}\cdot\nabla\tilde{\theta} 
    = - \nabla\cdot\boldsymbol{\pi}_\theta^{sgs};
\end{equation}
\begin{equation}
\label{eq:conc}
    \pdiff{\tilde{C}}{t}+\tilde{\boldsymbol u}\cdot\nabla\tilde{C} 
    = - \nabla\cdot\boldsymbol{\pi}_C^{sgs}.
\end{equation}
The tilde notation in \eqref{eq:mass}, \eqref{eq:mom}, \eqref{eq:heat}, 
and \eqref{eq:conc} represent the grid-filtered variables. 
In the Cartesian coordinate system $\boldsymbol{x} = (x,y,z)$, 
the velocity vector is $\tilde{\boldsymbol u}=(\tilde u,\tilde v,\tilde w)$, 
which represents the cross-front, along-front, and vertical components, respectively. 
The filtered potential temperature is $\tilde \theta$ in \eqref{eq:heat}, 
and the filtered tracer concentration is $\tilde C$ in \eqref{eq:conc}. 
The code has been extensively applied in previous studies of 
oceanic boundary layer flows 
\citep{chen2016,chor2018,yan2021,bo2024,bo2024nutrient,bo2025}. 

In \eqref{eq:mom}, $g$ is the gravitational acceleration and 
$\boldsymbol{e}_z$ is the unit vector in the vertical direction. 
The Coriolis frequency is $f=10^{-4}$~s$^{-1}$, corresponding 
to a latitude of around 45$^\circ$~N. 
The term $\tilde{p}^*$ denotes the modified pressure. 
The buoyancy variable is $\tilde b = g(\rho_0-\tilde\rho)/\rho_0$, 
where $\tilde{\rho}$ is the filtered density 
and $\rho_0$ is the reference density. 
Density variations are assumed to be only due to changes 
in potential temperature $\tilde\theta$ through 
a linear relationship $\tilde\rho=\rho_0[1-\alpha(\tilde\theta-\theta_0)]$. 
Here $\alpha=2\times10^{-4}$~K$^{-1}$ is the thermal 
expansion coefficient and $\theta_0$ is the reference 
potential temperature corresponding to $\rho_0$. 
The term $\boldsymbol{\tau}^{sgs}$ denotes the subgrid-scale 
(SGS) stress tensor, which is modeled using the Lagrangian 
scale-dependent dynamic Smagorinsky model \citep{bou2005}. 
The term $\boldsymbol{\pi}_\theta^{sgs}$ represents the 
SGS temperature flux. 
The SGS heat flux is modeled using the eddy diffusivity 
obtained from SGS viscosity and a prescribed value 
of turbulent Prandtl number $Pr_t = 0.4$. 
The term $\boldsymbol{\pi}_C^{sgs}$ represents the 
SGS tracer flux. 
Molecular viscosity and diffusivity are assumed to be negligible for 
high Reynolds number flows examined in this study. 
For simplicity, we will omit the tilde symbols used to 
denote grid-filtered variables elsewhere in the text. 

The flow field is initialized by adding an artificial relaxation 
damping term in the temperature equation, following a similar method 
to previous studies \citep{sullivan2018,sullivan2024,bo2025front}. 
The temperature field is forced toward the specified temperature 
field, i.e., the initial filament structure 
overlying the stratification. The surface heat flux is also applied 
during this period. Once the velocity field adjusts to the temperature 
field and both secondary circulation and convective turbulence develop, 
the relaxation term is removed. The removal of this forcing is defined 
as $t=0$, and after this point, the flow, temperature, and tracer 
concentration fields evolve freely in the LES. More details on the 
relaxation damping method can be found in \citet{sullivan2018}. 

The present LES framework resolves the flow and temperature 
fields using a grid structure with horizontally collocated 
points and a vertically staggered arrangement. 
A pseudospectral method is used in the 
horizontal direction, and vertical derivatives are discretized 
using a second-order central finite-difference method. 
The tracer transport equation is solved using a bounded 
finite-volume scheme \citep{chamecki2008,chor2018}. 
Periodic boundary conditions are applied in both horizontal 
directions, and vertical boundary conditions will be 
introduced in the following section. 
This study mainly focuses on submesoscale filaments with a 
symmetric configuration, while single-sided fronts are also 
considered for comparison. To simulate the horizontally 
asymmetric configuration of single-sided fronts under 
periodic boundary conditions, a temperature decomposition 
approach is used combined with an imposed baroclinic 
pressure gradient \citep{momen2018,bo2025front}. 

\subsection{Simulation setup}
The model domain spans 
$L_x\times L_y\times L_z=4000\times1000\times120$~m$^3$, 
discretized using $768\times192\times120$ uniformly spaced 
grid cells. 
The horizontal grid spacing is $\Delta x=\Delta y=5.2$~m, 
and the vertical grid spacing is $\Delta z=1$~m. 
The $x$-direction corresponds to the cross-front direction, 
and the $y$-direction to the along-front direction. 
At the upper boundary, a constant upward surface heat flux 
$Q$ (surface cooling) of 100~W~m$^{-2}$ is imposed. 
Surface wind stress is set to zero, and surface gravity waves 
are not included. We note that wind forcing and surface waves 
can modulate frontal turbulence 
\citep{mcwilliams2013,hamlington2014,sullivan2019}. 
Nevertheless, this study uses a simplified setup, with 
convective surface cooling as the only boundary forcing 
\citep{sullivan2024,bo2025front}, 
to focus on the fundamental mechanisms of tracer transport 
in submesoscale frontal zones. 
A sponge layer is applied in the bottom 20\% of the 
domain to prevent the reflection of internal gravity waves. 
The simulation setup is generally consistent with that 
in \citet{bo2025front}. 

The initial surface mixed layer depth $h_{0,OML}$ is $55$~m. 
A stably stratified layer is beneath the mixed layer, where 
the temperature gradient gradually increases to a uniform 
value of $\partial\theta/\partial z|_\infty=0.05$~$^\circ$C~m$^{-1}$. 
In our study, density variations are only due to changes 
in potential temperature $\theta$, and we do not 
consider the influence of salinity. 
A filament is initialized within the mixed layer as a dense 
anomaly, characterized by a temperature minimum at the center 
of the domain in the $x$-direction (cross-front direction). 
The maximum cross-front temperature difference is 
$\Delta_x \theta = 0.1$~$^\circ$C, and the initial filament width 
is approximately 1~km. The corresponding frontal strength 
is $M^2/f^2 = 50$, where $f$ is the Coriolis frequency and 
\begin{equation}
    \label{eq:M2-def}
    M^2=\max\left(\pdiff{\langle b\rangle_y}{x}\right)\Big\vert_{t=0,z=0} 
\end{equation}
with $\langle b\rangle_y$ representing the along-front 
averaged buoyancy. Here we focus on submesoscale filaments 
because their symmetric structure allows 
clearer diagnosis of tracer transport pathways, and 
filaments are also known to exhibit stronger surface 
convergence and vertical transport than single-sided fronts 
\citep{mcwilliams2009,bo2025front}. 
Tracer transport associated with single-sided fronts will be 
discussed later. 

The simulation is initialized with the specified filament 
shape using a relaxation damping method as introduced earlier. 
This approach allows us to investigate  
submesoscale frontal dynamics with boundary layer turbulence 
and TTW secondary circulation coexisting in the initialization. 
We uniformly release tracer at $t=0$ near the sea surface, 
with an initial concentration $C_0$ across the entire horizontal 
domain and vertically from $z=-10$~m to $z=0$~m. 
The tracer is subsequently transported passively by the 
evolving frontal secondary circulation and turbulence. 
The surface-released tracer serves as a proxy for a range 
of materials concentrated 
near the surface, such as pollutants \citep{taylor2023}, 
dissolved gases \citep{follows1996,pietri2025}, or biological 
tracers like phytoplankton and particulate organic carbon 
\citep{omand2015,mahadevan2016}. 

\begin{table}
\begin{center}
\setlength{\tabcolsep}{10pt}
\begin{tabular}{ l l c l } 
\hline
Case & Frontal configuration & $h_{0,OML}$ [m] & Tracer release \\
\hline
Frontal $\star$ & Filament & 55 & Surface, domain-wide \\ 
Post-frontal & Decayed filament & 55 & Surface, domain-wide \\ 
No-front & Uniform mixed layer & 55 & Surface, domain-wide \\[8pt]
Deep frontal (SD) & Filament & 95 & Surface, domain-wide \\ 
Deep frontal (SF) & Filament & 95 & Surface, far-field \\ 
Deep frontal (MD) & Filament & 95 & Mixed layer, domain-wide \\ 
Deep frontal (MF) & Filament & 95 & Mixed layer, far-field \\ 
Deep no-front (SD) & Uniform mixed layer & 95 & Surface, domain-wide \\ 
Deep no-front (MD) & Uniform mixed layer & 95 &  Mixed layer, domain-wide \\[8pt] 
Deep frontal (I) & Filament & 95 & Interior, domain-wide \\[8pt]
Single-sided & Single-sided front & 55 & Surface, domain-wide \\ 
\hline
\end{tabular}
\caption{
Summary of numerical simulations. The baseline case 
selected for detailed analysis is marked with a star. 
The post-frontal case is initialized from the final 
state of the baseline frontal case. The no-front case 
features a uniform mixed layer without frontal 
structures. The single-sided case replaces the filament 
(two-sided front) with a single-sided front. 
The frontal, post-frontal, no-front, and single-sided 
cases all use the shallow mixed layer with near-surface, 
domain-wide tracer release. 
The deep cases have an increased mixed-layer depth, and 
the letters in parentheses denote the tracer-release 
configuration. Vertically, `S' indicates release near 
the sea surface and `M' indicates release throughout 
the depth of the mixed layer. 
Horizontally, `D' denotes release across the entire 
domain, whereas `F' denotes release in the far field. 
The letter `I' indicates release from the ocean interior.}
\label{tab:models}
\end{center}
\end{table}

The simulation described above is the baseline case for 
detailed analysis in this study and will henceforth be 
referred to as the frontal case. The following results 
and analysis will first focus on this baseline frontal 
case, while additional simulations are presented later 
for comparison. These additional simulations are conducted 
to examine the influences of frontal structure, mixed-layer 
depth, and tracer release configuration (Table~\ref{tab:models}). 
Specifically, we consider a post-frontal case initialized 
from the final state of the baseline simulation after the 
filament has substantially weakened. 
We also consider a no-front case, which has an otherwise 
identical setup to the frontal case, except that no 
frontal structure is present, i.e., zero horizontal 
temperature gradients in the ocean mixed layer ($M^2 = 0$). 
In addition, a simulation initialized with a single-sided 
front is conducted, which features a single density step 
rather than a symmetric filament which can be viewed as 
a two-sided front. 

Moreover, we conduct simulations of the submesoscale 
filament with an increased initial mixed layer depth 
$h_{0,OML}$ of 95~m. These deep mixed 
layer simulations are performed in a model domain of 
$L_x\times L_y\times L_z = 4000\times1000\times160$~m$^3$, 
discretized using $768\times192\times160$ uniformly 
spaced grid cells. The grid resolution, initial frontal 
strength, and surface boundary forcing are identical to 
those in the baseline frontal case. The only differences 
are the increased mixed layer depth and the 
correspondingly greater domain height. 
A series of tracer release experiments is conducted for 
the deep mixed layer configuration (Table~\ref{tab:models}, 
deep frontal cases SD, SF, MD, MF). Horizontally, tracer 
is initially released either across the entire domain or 
only in the far field, i.e., away from the filament 
center ($|x|>700$~m). Vertically, tracer is released 
either near the surface ($z=-10$~m to $z=0$~m, 
referred to as near-surface release) or throughout most 
of the mixed layer ($z=-80$~m to $z=0$~m, mixed-layer 
release). In addition, the no-front case ($M^2 = 0$) is 
examined for the deep mixed layer configuration for 
comparison (cases deep no-front SD and MD).

\subsection{Theoretical framework}
In this subsection, we describe the theoretical framework 
used to diagnose tracer transport in submesoscale frontal 
zones, with a particular focus on diapycnal transport. 
The velocity field can be decomposed into mean flow and temporal 
fluctuation components, 
\begin{equation}\label{eq:flow-decomp}
    \boldsymbol{u} = \overline{\boldsymbol{u}} + \boldsymbol{u}'.
\end{equation}
The overline denotes the time average and the prime denotes 
fluctuations relative to the average, with a time-averaging 
window of 150~s. 
This decomposition separates two distinct time scales, 
where $\boldsymbol{u}'$ primarily represents turbulent 
motions, and $\overline{\boldsymbol{u}}$ represents more 
slowly evolving structures such as secondary circulation. 
Similarly, the tracer concentration is decomposed as 
\begin{equation}\label{eq:conc-decomp}
    C = \overline{C} + C',
\end{equation}
where $\overline{C}$ is the slowly evolving mean 
concentration and $C'$ denotes rapidly varying 
turbulent fluctuations. 

In a fixed control volume $C\V$, the time-averaged 
mass conservation of any scalar, such as the tracer
concentration ${C}$ considered in the present study, 
is given by 
\begin{equation}
    \label{eq:CV}
   \int_{C\V} \pdiff{\overline C}{t}\infd \V + \int_{CS} 
   (\boldsymbol{\overline u}\overline C)\cdot\hat{n}\infd A + \int_{CS} \overline{\boldsymbol{u'}C'}\cdot\hat{n}\infd A = 0, 
\end{equation}
Here $CS$ denotes the control surface enclosing $C\V$, 
and $\hat{n}$ is the outward unit vector normal to $CS$. 
Note that this is a windowed time average taken over a 
period much longer than the timescale of turbulent 
fluctuations, and the averaged field can still evolve in time. 
The second and third terms represent the mean advective and 
turbulent fluxes across the control surface $CS$, respectively. 
No internal sources or sinks are considered here. 
Expressed in differential form, equation~\eqref{eq:CV} becomes 
\begin{equation}
    \label{eq:mass-2}
    \pdiff{\overline C}{t} + \boldsymbol{\overline u}\cdot\nabla\overline C + \nabla\cdot\overline{\boldsymbol{u'}C'} = 0,
\end{equation}
where we have used the incompressibility condition 
$\nabla\cdot\boldsymbol{\overline u}=0$. 

In a deformable control volume $C\V^*$, the time-averaged mass 
conservation is \citep[e.g.,][]{kundu2024} 
\begin{equation}
    \label{eq:CV-iso}
   \diff{}{t}\int_{C\V^*}\overline C\infd \V^* + 
   \int_{CS^*} (\boldsymbol{\overline u}_r \overline C)\cdot\hat{n}^*\infd A^* 
   + \int_{CS^*} \overline{\boldsymbol{u'}C'}\cdot\hat{n}^*\infd A^* =0.
\end{equation}
where $CS^*$ denotes the moving boundaries (control surface) of $C\V^*$, 
and $\hat{n}^*$ is the outward unit normal vector. 
The relative velocity between the fluid and moving boundaries 
is $\boldsymbol{\overline u}_r$. 
In this study, the moving boundaries are chosen to follow 
the slowly evolving mean isopycnals, calculated from the 
windowed time average in \eqref{eq:conc-decomp}. 
The definition of mean isopycnals filters out instantaneous 
turbulent fluctuations, and this timescale separation ensures 
the control volume boundaries evolve smoothly enough for the 
conservation law to hold. Accordingly, the relative velocity 
can be written as $\boldsymbol{\overline u}_r= 
\boldsymbol{\overline u}-\boldsymbol{\overline u}_{iso}$, 
where $\boldsymbol{\overline u}_{iso}$ is 
the mean velocity of isopycnals. 
In such case, the second and third terms represent the mean 
advective and turbulent diapycnal fluxes across the control 
surface $CS^*$ (moving isopycnals), respectively. 
We define the total mean diapycnal velocity as 
\begin{equation}
    \label{eq:vperp}
    V_\perp=\boldsymbol{\overline u}_r\cdot\hat{n}^*, 
\end{equation}
which quantifies the net motion of fluid parcels across isopycnals.
Therefore, the total advective diapycnal flux can be expressed as
\begin{equation}
    \label{eq:vperp-flux}
    (\boldsymbol{\overline u}_r \overline C)\cdot\hat{n}^*=V_\perp \overline C. 
\end{equation}
According to the Leibniz theorem \citep{kundu2024}, 
\begin{equation}
    \label{eq:CV-iso-sep}
   \diff{}{t}\int_{C\V^*}\overline C\infd \V^* = \int_{C\V^*} \pdiff{\overline C}{t}\infd \V^* + 
   \int_{CS^*} (\boldsymbol{\overline u}_{iso}\overline C)\cdot\hat{n}^*\infd A^*, 
\end{equation}
so equation~\eqref{eq:CV-iso} can be rewritten as 
\begin{equation}
    \label{eq:CV-iso2}
   \int_{C\V^*} \pdiff{\overline C}{t}\infd \V^* + 
   \int_{CS^*} (\boldsymbol{\overline u}_{iso}\overline C)\cdot\hat{n}^*\infd A^*
   + \int_{CS^*} (\boldsymbol{\overline u}_r\overline C)\cdot\hat{n}^*\infd A^* 
   + \int_{CS^*} \overline{\boldsymbol{u'}C'}\cdot\hat{n}^*\infd A^*=0.
\end{equation}
Expressed in differential form, 
\begin{equation}
    \label{eq:mass-iso-2}
    \pdiff{\overline C}{t} + 
    \boldsymbol{\overline u}_{iso}\cdot\nabla\overline C 
    + \boldsymbol{\overline u}_r\cdot\nabla\overline C
    + \nabla\cdot\overline{\boldsymbol{u'}C'} = 0, 
\end{equation}
where the incompressible condition is used again. 
While equation~\eqref{eq:mass-iso-2} is mathematically 
equivalent to equation~\eqref{eq:mass-2}, we present it 
in this form to explicitly isolate the contribution from 
diapycnal transport. 

Density depends only on temperature in this study, so 
isopycnals coincide with isotherms and diapycnal transport 
corresponds to diathermal transport. For consistency, we 
retain the terms isopycnals and diapycnal transport. 
When the scalar $C$ is potential temperature $\theta$, 
we obtain 
\begin{equation}
    \label{eq:temp-iso-2}
    \pdiff{\overline \theta}{t} + 
    \boldsymbol{\overline u}_{iso}\cdot\nabla\overline \theta=0, 
\end{equation}
because temporal changes in local temperature are directly 
related to the movement of isopycnals; in other words, 
temperature is conserved when following isopycnals. 
Combining this with equation~\eqref{eq:mass-iso-2} yields a 
direct relationship between the mean diapycnal velocity and 
the turbulent flux 
\begin{equation}
    \label{eq:temp-iso-3}
    \boldsymbol{\overline u}_r\cdot\nabla\overline \theta
    = V_\perp |\nabla\overline \theta|
    = -\nabla\cdot\overline{\boldsymbol{u'}\theta'}.
\end{equation}
In the one-dimensional case where only vertical transport is 
considered, equation~\eqref{eq:temp-iso-3} reduces to the 
classical formulation \citep{kullenberg1977,mcdougall1990,fernando1991,wang2017} 
\begin{equation}
    \label{eq:temp-entrainment}
    V_\perp = \frac{-\partial(\overline{{w'}\theta'})/\partial z}{\partial \theta/\partial z} 
    \sim \frac{-\overline{{w'}\theta'}}{\Delta \theta}.
\end{equation}
In this context, the mean diapycnal velocity $V_\perp$ is often 
referred to as the entrainment velocity at the interface between 
the turbulent boundary layer and stratified interior. 
Equation~\eqref{eq:temp-iso-3} reveals the fundamental physics 
behind the mean diapycnal velocity: turbulent mixing of density 
creates a relative motion between the fluid and isopycnals, 
which can subsequently lead to an advective diapycnal flux. 
It is important to note that equation~\eqref{eq:temp-iso-3} 
applies only to density (here a function of temperature alone). 
For other scalar tracers, the balance between the first 
two terms in equation~\eqref{eq:mass-iso-2}, i.e., the time 
derivative and advection by $\boldsymbol{\overline u}_{iso}$, 
does not generally hold. As a result, although advective diapycnal 
transport (the third term in equation~\eqref{eq:mass-iso-2}) 
can occur when a mean diapycnal velocity is present, it is 
not necessarily comparable in magnitude to turbulent transport 
(the last term in equation~\eqref{eq:mass-iso-2}). 

\section{Results and analysis}
\label{sec:results}

\subsection{Frontal dynamics}
\label{sec:results-front}
\begin{figure}
\centering
  \includegraphics[width=1\textwidth]{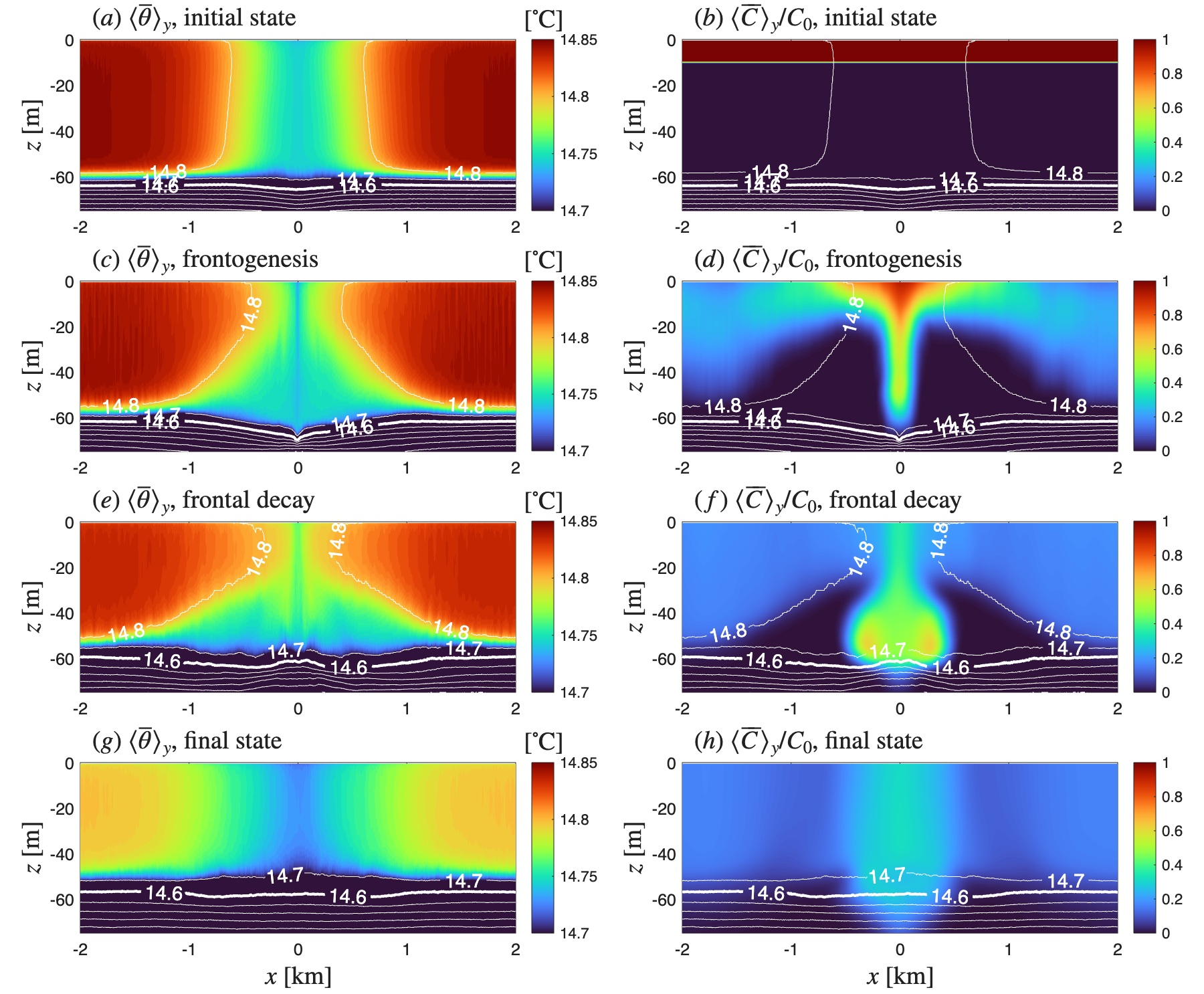}
  \caption{Side views of the along-front averaged temperature 
  (left column) and tracer concentration (right column) fields. 
  $(a)$, $(c)$, $(e)$, and $(g)$: Temperature fields at four 
  representative times: the initial state ($t=0$), the 
  frontogenesis phase ($t=1.5$~hr), the early stage of frontal 
  decay ($t=5$~hr), and the final state ($t=20$~hr), respectively. 
  $(b)$, $(d)$, $(f)$, and $(h)$: Tracer distributions at the 
  same four times, overlaid with temperature contours. 
  Tracer concentration is normalized by the initial concentration $C_0$.}
\label{fig:temp}
\end{figure}

A key mechanism driving submesoscale frontogenesis is the 
turbulent thermal wind (TTW), which arises from an approximate 
momentum balance among the pressure gradient, Coriolis 
force, and the vertical divergence of turbulent momentum 
flux \citep[e.g.,][]{mcwilliams2016,sullivan2018}. 
TTW gives rise to a pair of counter-rotating secondary 
circulation cells centered around the filament in the 
cross-front section (figure~\ref{fig:vel}$a$). 
These secondary circulation cells converge near the 
surface, thereby enhancing frontogenesis and 
sharpening the gradients of momentum and temperature 
at the filament center (near $x=0$, see figure~\ref{fig:temp}$c$). 
The surface convergence also drives enhanced downwelling 
during frontogenesis (figure~\ref{fig:vel}$b$). 
It is worthwhile to mention that 
the submesoscale filament sharpens to widths of less 
than 100~m during frontogenesis, which is below the 
grid resolution of many ocean models. 

Various instabilities can develop as frontal gradients intensify, 
leading to the growth of frontal turbulence 
(figure~\ref{fig:vel}$c$). When the growth rate 
of instabilities exceeds the frontogenetic rate, cross-front 
turbulent fluxes act to halt further sharpening, resulting in 
frontal arrest. After reaching peak frontal strength at arrest, 
the filament enters an early decay stage 
(figure~\ref{fig:temp}$e$), when the frontal gradients 
of momentum and temperature gradually weaken. 
Turbulence intensity in the frontal zone remains elevated 
compared to background levels in the mixed layer, though 
lower than its peak values around the time of frontal arrest. 
This is followed by a late decay stage throughout which the 
frontal gradients remain relatively steady, and a weak 
filament structure, along with secondary circulation and 
downwelling, persists until the end of the simulation. 

Turbulence plays a critical role throughout the evolution of the 
filament, from its initial intensification to the eventual decay. 
The interaction between turbulence and frontal dynamics has 
been examined in detail in previous studies 
\citep{sullivan2018,sullivan2024,bo2025front}. 
Here we shift the focus to how turbulence affects tracer 
transport at the submesoscale filament. 
It is evident that tracer penetrates below the initial mixed 
layer depth in the frontal zone, where intense vertical motions 
occur (figure~\ref{fig:temp}$b$, $d$, $f$). 
In addition, frontogenesis is typically accompanied by 
restratification in the surface mixed layer 
\citep[e.g.,][]{freilich2021,taylor2023}. 
Secondary circulation induces frontal slumping, with 
denser water displaced downward and lighter water upward 
on both flanks of the filament (figure~\ref{fig:temp}$c$). 
Nevertheless, despite restratification, tracer is transported 
to depths exceeding the initial mixed layer depth, further 
highlighting the importance of submesoscale processes in 
enhancing vertical transport into the ocean interior. 
The physical mechanisms driving this tracer transport 
will be investigated in the following sections.

\subsection{Turbulence and diapycnal velocity}
\label{sec:results-dia-vel}
\begin{figure}
\centering
  \includegraphics[width=1\textwidth]{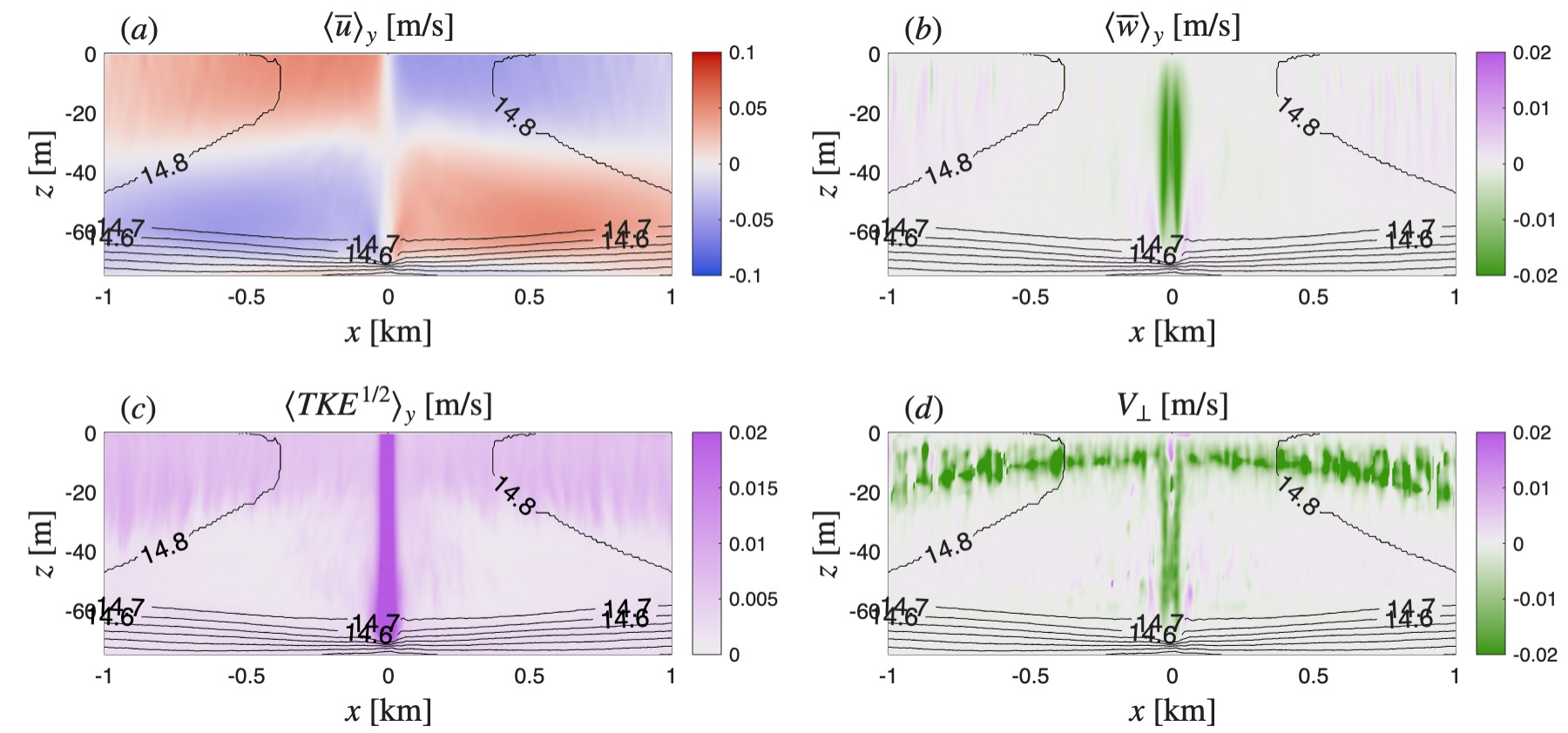}
  \caption{Side views of mean and turbulent velocity fields 
  around peak frontal strength. Results are averaged in the 
  along-front direction and overlaid with temperature contours.
  $(a)$: Mean cross-front velocity. 
  $(b)$: Mean vertical velocity. 
  $(c)$: Square root of the turbulent kinetic energy, representing a turbulent 
  velocity scale. 
  $(d)$: Mean diapycnal velocity, i.e., the projection of the relative velocity 
  onto the local isopycnal-normal direction, as defined by equation~\eqref{eq:vperp}. 
  Negative values indicate flow toward lower-temperature regions.}
\label{fig:vel}
\end{figure}

Downwelling associated with submesoscale fronts is a widely 
recognized pathway for tracer transport. 
However, it remains unclear whether this advective transport
occurs predominantly along isopycnals or whether it also 
involves a significant diapycnal component. 
This distinction is important, as diapycnal transport implies 
more effective material exchange between water masses with
distinct physical and biogeochemical properties \citep{ledwell1998}. 
The intense turbulence generated in the frontal zone 
(figure~\ref{fig:vel}$c$) raises the possibility of enhanced 
diapycnal transport. 
We examine diapycnal motion by calculating the time-averaged 
relative velocity $\boldsymbol{\overline u}_r=
\boldsymbol{\overline u}-\boldsymbol{\overline u}_{iso}$, 
where the velocity of isopycnals $\boldsymbol{\overline u}_{iso}$ 
is determined by tracking the displacement of mean isopycnals 
between consecutive time steps of 150~s and is meaningful 
only in the direction normal to the isopycnals. 
This velocity difference between fluid parcels and 
isopycnals leads to a strong mean diapycnal velocity 
(figure~\ref{fig:vel}$d$), calculated based on 
equation~\eqref{eq:vperp}. 

In classical studies of boundary layer entrainment 
\citep[e.g.,][]{mcdougall1990,fernando1991}, the diapycnal 
velocity, often referred to as the entrainment velocity 
in that context, reflects the growth rate of boundary 
layer thickness. In such cases, there is no time-averaged 
vertical velocity of fluid parcels; rather, turbulent mixing displaces 
isopycnals, creating a relative motion between fluid and 
isopycnals. In contrast, at the submesoscale filament, the 
dominant contribution to diapycnal velocity arises from 
the secondary circulation and associated downwelling. 
In our study, strong time-averaged fluid motions are present, 
but intense turbulent mixing prevents isopycnals from 
following the fluid parcels adiabatically. This causes the 
isopycnals to lag behind the fluid motion, as seen in the 
predominantly negative $V_\perp$ in figure~\ref{fig:vel}$(d)$, 
resulting in a net diapycnal transport. 
The mean diapycnal velocity can be estimated either 
from the relative motion between fluid and isopycnals 
(equation~\eqref{eq:vperp}) or from turbulent 
density mixing (equation~\eqref{eq:temp-iso-3}), 
both yielding consistent results. 

In addition to the mean diapycnal velocity, which can drive 
advective transport, frontal turbulence can also directly 
contribute to tracer transport through the turbulent flux 
(eddy covariance). 
We quantify turbulence intensity using the turbulent kinetic 
energy (TKE), defined as 
\begin{equation}
    \label{eq:tke-component}
    \textrm{TKE} = \frac{1}{2}\left( \overline{u'^2}+\overline{v'^2}+\overline{w'^2} \right). 
\end{equation}
It is worth noting that the magnitude of the mean 
diapycnal velocity (figure~\ref{fig:vel}$d$) is comparable 
to the turbulent velocity (figure~\ref{fig:vel}$c$). 
Although we focus on a single snapshot 
near the time of frontal arrest in this analysis, the 
mean diapycnal velocity persists for several hours from 
frontogenesis through the early frontal decay stage, during which 
turbulence intensity also remains elevated in the frontal 
zone compared to background levels. 

\subsection{Tracer transport}
\label{sec:results-dia-flux}
As illustrated in figure~\ref{fig:temp}, the passive 
tracer released near the sea surface is transported to 
the ocean interior in the submesoscale frontal zone. 
To further assess the influence of frontogenesis in 
driving diapycnal transport, we compare the baseline 
frontal case with two reference simulations -- 
the post-frontal case and the no-front case. 

In the frontal case, passive tracer is released before 
frontogenesis, which subsequently drives strong 
diapycnal transport. The post-frontal case is initialized 
from the final state of the frontal case, 
with the tracer released at its initialization, and run for 
the same duration. A weak, decayed filament and residual 
downwelling still exist in the post-frontal case, as in the 
late decay stage in the frontal case, but it lacks strong 
frontogenesis as well as the rapid growth of frontal 
instability and turbulence. The post-frontal case can therefore 
serve as a reference simulation, with the tracer released 
after frontal decay and not experiencing frontogenesis. 

\begin{figure}
\centering
  \includegraphics[width=\textwidth]{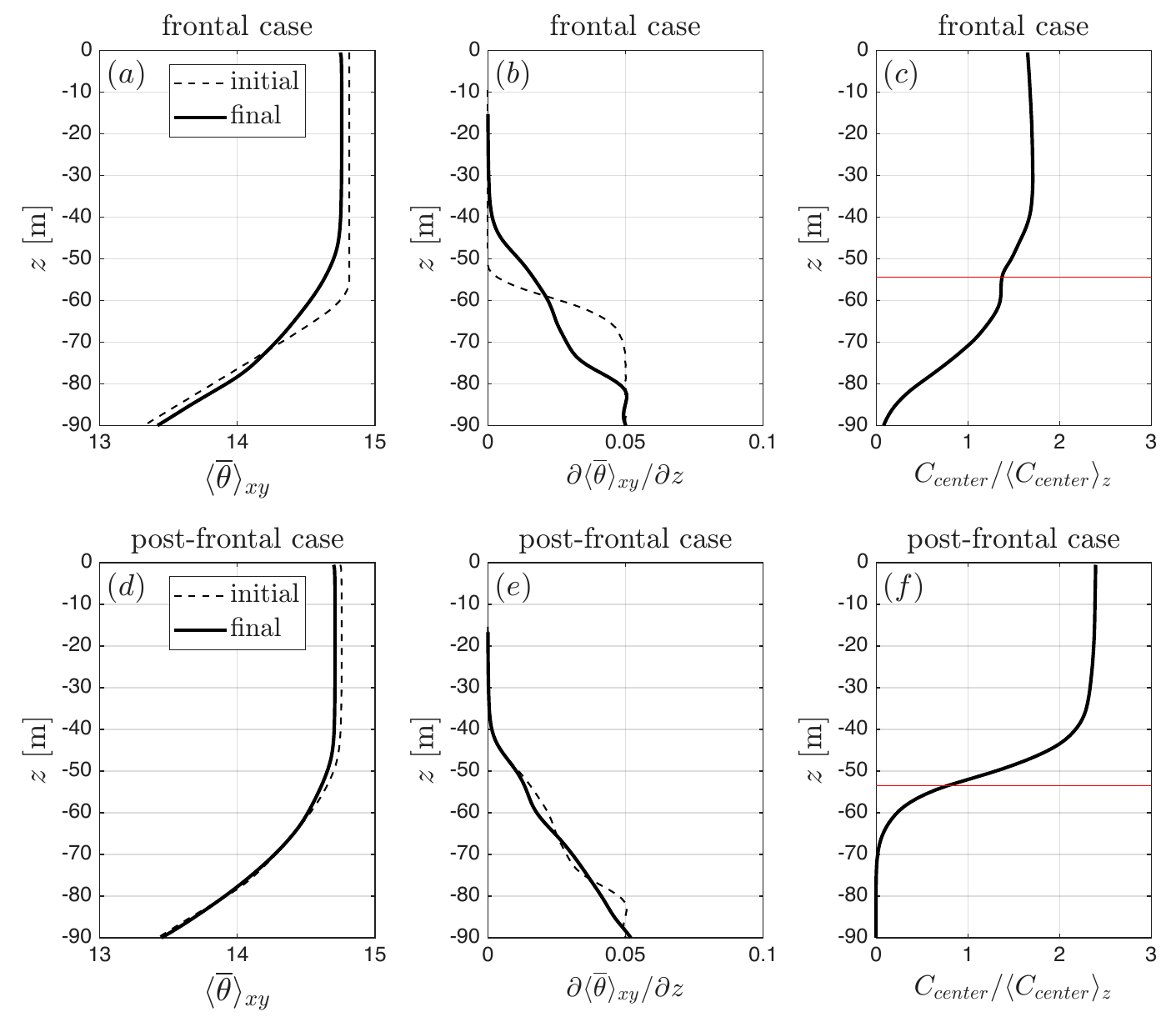}
  \caption{Vertical profiles of temperature and tracer concentration 
  in the frontal and post-frontal cases. 
  $(a)-(c)$ Frontal case. $(d)-(f)$ Post-frontal case. 
  $(a)$ and $(d)$: Horizontally averaged temperature. The dashed line shows 
  the initial state, and the solid line shows the final state. 
  $(b)$ and $(e)$: Horizontally averaged vertical temperature gradient. 
  $(c)$ and $(f)$: Vertical distribution of tracer concentration at the 
  filament center in the final state. 
  Here $C_{center}/\langle C_{center}\rangle_z $ denotes 
  the along-front averaged tracer concentration at $x=0$, 
  normalized its depth average. The horizontal red line 
  indicates the depth of the 14.6~$^\circ$C isotherm.}
\label{fig:vert-prof-front}
\end{figure}

In the frontal case, frontogenesis leads to restratification 
of the surface mixed layer (figure~\ref{fig:temp}$c$), 
resulting in a shallower mixed layer depth 
(figure~\ref{fig:vert-prof-front}$a$). 
The mixed layer depth is typically defined using a 
temperature deviation of 0.2~$^\circ$C from the sea 
surface \citep{de2004}; in our setup the base of the 
mixed layer approximately corresponds to the 
14.6~$^\circ$C isotherm. 
Alternatively, the mixed layer depth may be defined 
using a threshold of small vertical temperature gradients. 
Using different definitions does not change the conclusion 
that frontogenesis shallows the mixed layer, a phenomenon 
widely supported by observational evidence \citep[e.g.,][]{taylor2023}. 
Despite this restratification, the tracer still penetrates 
to depths of 80~m (figure~\ref{fig:vert-prof-front}$c$), 
which is even greater than the original mixed layer depth. 
It is worth noting that a transitional 
region forms below the mixed layer (e.g., from 40~m to 80~m), 
bounded above by the depth where the vertical temperature 
gradient reaches $\partial\theta/\partial z|_\infty$. 
This region reflects the vertical extent influenced by 
frontogenesis and the associated diapycnal mixing. 
Although this region is located below the mixed layer 
depth after frontogenesis, it retains a memory of 
previous disturbances by submesoscale frontal processes. 
Consequently, it is expected that the tracer penetration 
depth aligns with the depth of this transitional region. 

In contrast, the post-frontal case exhibits relatively steady 
temperature profiles (figure~\ref{fig:vert-prof-front} $d$-$f$), 
with only minor changes due to surface cooling and boundary 
layer entrainment. The tracer, released after frontal decay, 
largely remains within the mixed layer. Only a small portion 
reaches depths of 60~m, primarily due to boundary layer 
mixing associated with convective turbulence. 
This comparison highlights the enhanced tracer transport into 
the ocean interior in the frontal compared to the post-frontal 
case, indicating the important role of submesoscale 
frontogenesis and associated turbulence. 

\begin{figure}
\centering
  \includegraphics[width=\textwidth]{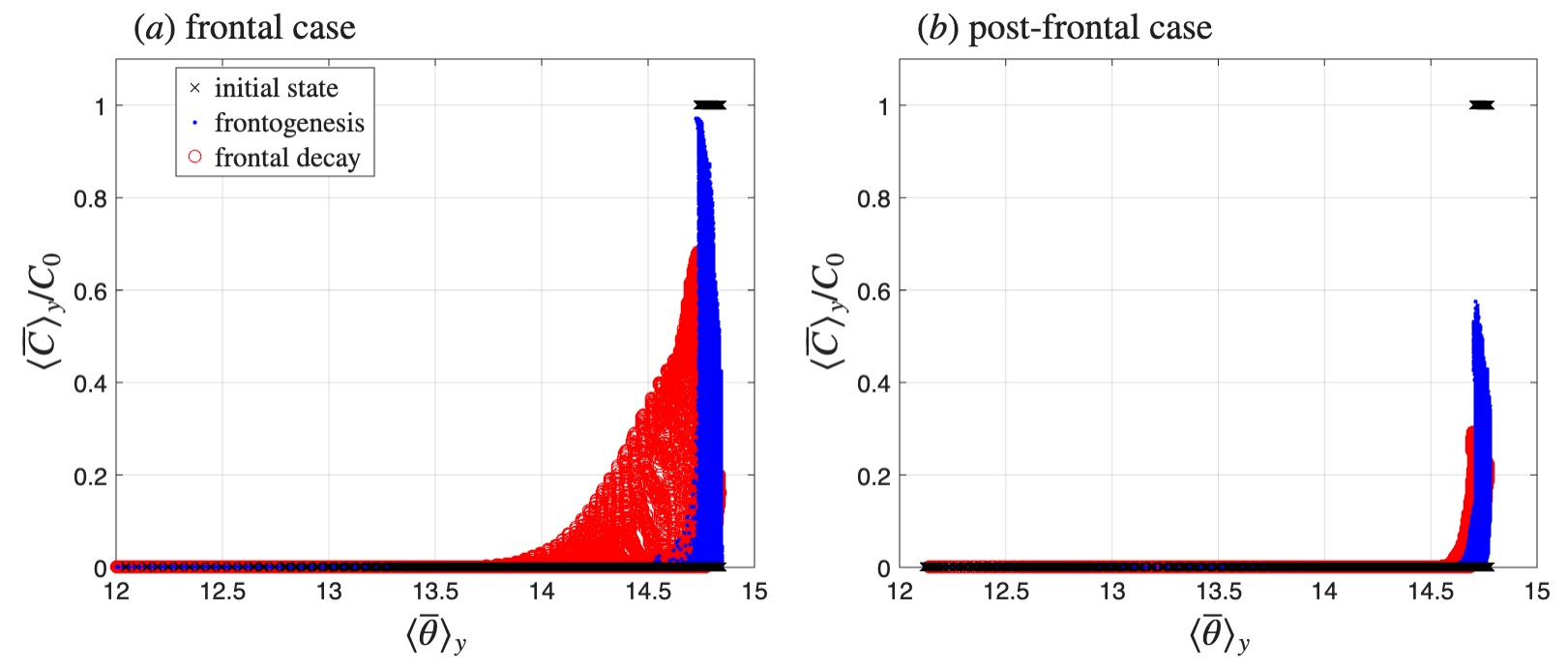}
  \caption{Tracer concentration–temperature diagram. 
  $(a)$ Frontal case. Three representative times are shown: 
  the initial state, a time during the frontogenesis stage, 
  and a time during the frontal decay stage. 
  $(b)$ Post-frontal case. The same three times corresponding 
  to panel (a) are shown. See the legend for details.}
\label{fig:Pcon_temp_corr}
\end{figure}

Diapycnal transport is also evident in the tracer 
concentration–temperature diagram 
(figure~\ref{fig:Pcon_temp_corr}$a$). 
In the frontal case, the tracer initially released in the 
surface mixed layer at higher temperatures rapidly spreads 
across a broad range in the temperature coordinate due to 
frontogenesis and diapycnal fluxes. This contrasts with the 
classical view of subduction, where submesoscale frontal 
downwelling is assumed to occur predominantly along 
isopycnals \citep[e.g.,][]{qu2022,taylor2023}. 
Our study suggests that downwelling can also drive advective 
diapycnal transport, revealing an underexplored mechanism 
that presents new challenges for modeling and observing 
tracer transport. 
As a comparison, the tracer concentration–temperature 
diagram (figure~\ref{fig:Pcon_temp_corr}$b$) reflects 
weaker diapycnal transport in the post-frontal case, 
where high tracer concentrations are mostly confined 
within a narrow temperature range. 

\begin{figure}
\centering
  \includegraphics[width=\textwidth]{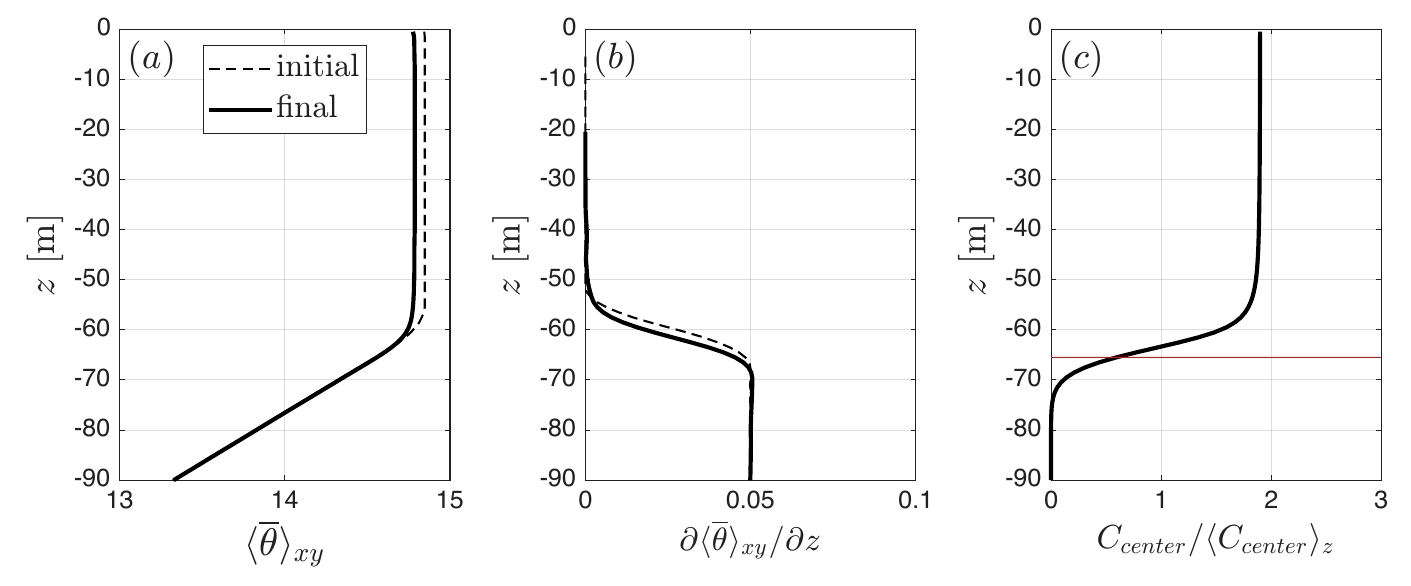}
  \caption{Vertical profiles of temperature and tracer concentration 
  in the no-front case. 
  $(a)$: Horizontally averaged temperature. The dashed line shows 
  the initial state, and the solid line shows the final state. 
  $(b)$: Horizontally averaged vertical temperature gradient. 
  $(c)$: Vertical distribution of tracer concentration at the 
  domain center in the final state. The horizontal red line 
  indicates the depth of the 14.6~$^\circ$C isotherm.}
\label{fig:vert-prof-no-front}
\end{figure}

In the no-front case, the tracer is released in a horizontally 
uniform ocean mixed layer without any frontal structures. 
The mixed layer depth remains relatively unchanged 
throughout the simulation (figure~\ref{fig:vert-prof-no-front}), 
compared to the restratified mixed layer in the frontal case. 
Tracer transport is driven solely by convective turbulence 
associated with sea surface cooling, with only a small 
fraction of tracer penetrating into the ocean interior. 

\begin{figure}
\centering
  \includegraphics[width=0.6\textwidth]{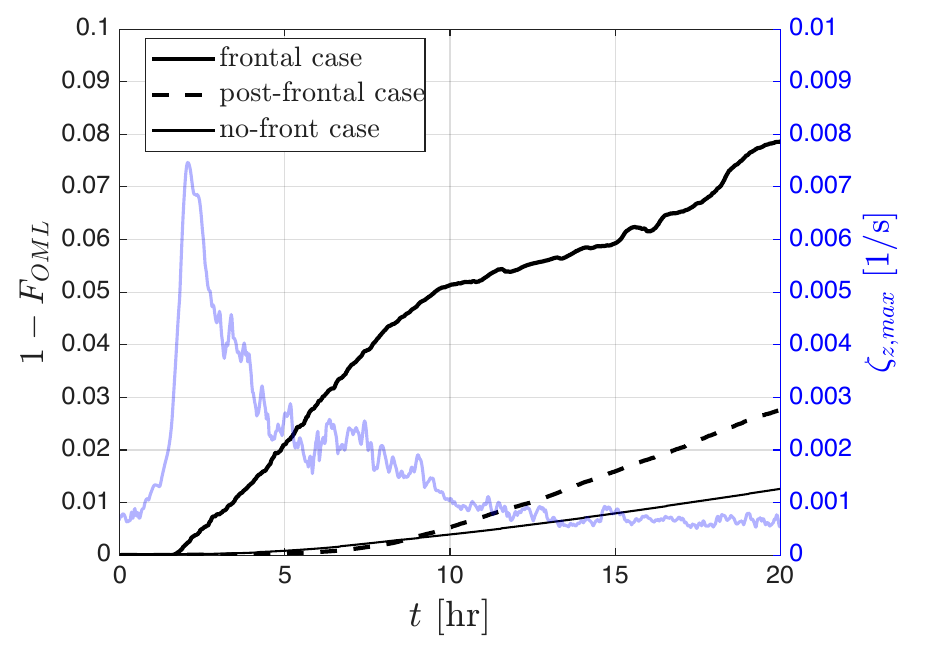}
  \caption{ Time series of $1-F_{OML}$, the fraction of tracer 
  that has penetrated below the ocean mixed layer. 
  Solid black line: frontal case (baseline simulation with 
  tracer released before frontogenesis). 
  Dashed black line: post-frontal case (reference simulation with 
  tracer released after frontal decay, not experiencing frontogenesis). 
  Dotted black line: no-front case (reference simulation with 
  tracer released in a horizontally uniform ocean mixed layer). 
  Blue line (right axis): time series of frontal strength in the frontal case, 
  quantified as the maximum along-front averaged vertical vorticity $\zeta_{z,max}$.
  }
\label{fig:Pcon-time}
\end{figure}

Finally, we compare the frontal, post-frontal, and no-front cases 
and quantify the fraction of tracer transported into 
the ocean interior as $1-F_{OML}$. Here $F_{OML}$ 
denotes the fraction of tracer remaining in the ocean 
mixed layer, and 
\begin{equation}
    \label{eq:tracer-fraction}
    1-F_{OML} = 1-\frac{\int_{L_x}\int_{L_y}\int_{-h_{OML}}^0 C \infd x\infd y\infd z}
    { \V_0 C_0} = \frac{\int_{L_x}\int_{L_y}\int_{-L_z}^{-h_{OML}} C \infd x\infd y\infd z}
    { \V_0 C_0}. 
\end{equation}
In this equation, $\V_0$ is the volume over which the 
tracer is initially released, and 
$h_{OML}$ is the ocean mixed layer depth defined by the 
14.6~$^\circ$C isotherm \citep{de2004}. 
Both the frontal and post-frontal cases exhibit a greater 
fraction of tracer penetrating into the stratified 
interior compared to the no-front case, highlighting the 
role of submesoscale processes in enhancing diapycnal 
transport. Moreover, in the frontal case, approximately 
three times more tracer is transported beneath  
the mixed layer compared to the post-frontal case 
(figure~\ref{fig:Pcon-time}). 
This underscores the importance of resolving turbulent 
frontogenesis and its associated diapycnal tracer transport. 
Yet these processes are likely underrepresented in many 
ocean models, as discussed later. 
Notably, rapid tracer transport begins around $t=1$~hr, 
corresponding to the onset of frontogenesis, and continues 
through the early decay stage until $t=10$~hr 
(figure~\ref{fig:Pcon-flux}), during which frontal 
gradients and turbulence remain relatively strong. 
After $t=10$~hr, the tracer transport rate decreases 
as the filament transitions into the late decay stage, becoming  
comparable to that in the post-frontal case. This is expected 
since the post-frontal case effectively represents an 
extension of the late decay stage, with both characterized 
by a weak filament structure. 

We note that the frontal case exhibits a lower initial 
temperature at the filament center than the no-front case. 
As a result, tracer released near the sea surface can more 
readily cross a given isotherm (e.g., the 14.6~$^\circ$C 
isotherm near the base of the mixed layer) in the frontal 
case than in the no-front case. 
To address this potential bias, we conduct an additional 
no-front simulation initialized with a reduced mixed-layer 
temperature, matching the minimum sea surface temperature 
in the frontal case, i.e., the temperature at the filament 
center. In this setup, a larger fraction of tracer is 
initially located at lower temperatures, making it more 
favorable for tracer to cross the 14.6~$^\circ$C isotherm 
into the ocean interior. Nevertheless, the amount of tracer 
transported below the mixed layer is still substantially 
smaller than in the frontal case. 

It is also worth mentioning that tracer concentration 
exhibits cross-front variability (figure~\ref{fig:temp}). 
Submesoscale fronts can modify not only vertical tracer 
transport but also the horizontal distribution of tracer, 
such as enhanced horizontal mixing or localized tracer 
accumulation associated with frontal convergence 
\citep[e.g.,][]{taylor2018,wenegrat2020}. 
However, horizontal transport associated with frontogenesis 
is not the focus of the present study, which emphasizes 
vertical transport. In figures~\ref{fig:vert-prof-front} 
and \ref{fig:vert-prof-no-front}, 
horizontally averaged temperature profiles are shown because 
temperature is relatively more uniform in the horizontal 
direction, whereas tracer profiles are shown at the 
filament center where frontal tracer transport is strongest. 
The tracer concentration profile is 
normalized by its local depth-averaged value rather than by 
its domain-averaged value. The depth-averaged tracer 
concentration is higher at the filament center than in regions 
away from the frontal zone. Therefore, if the normalization 
were instead performed using the domain-averaged tracer 
concentration, it would further increase the apparent 
relative tracer concentration at the filament center, 
including that penetrating below the mixed layer 
(figure~\ref{fig:vert-prof-front}$c$).

\subsection{Diapycnal fluxes}

\begin{figure}
\centering
  \includegraphics[width=\textwidth]{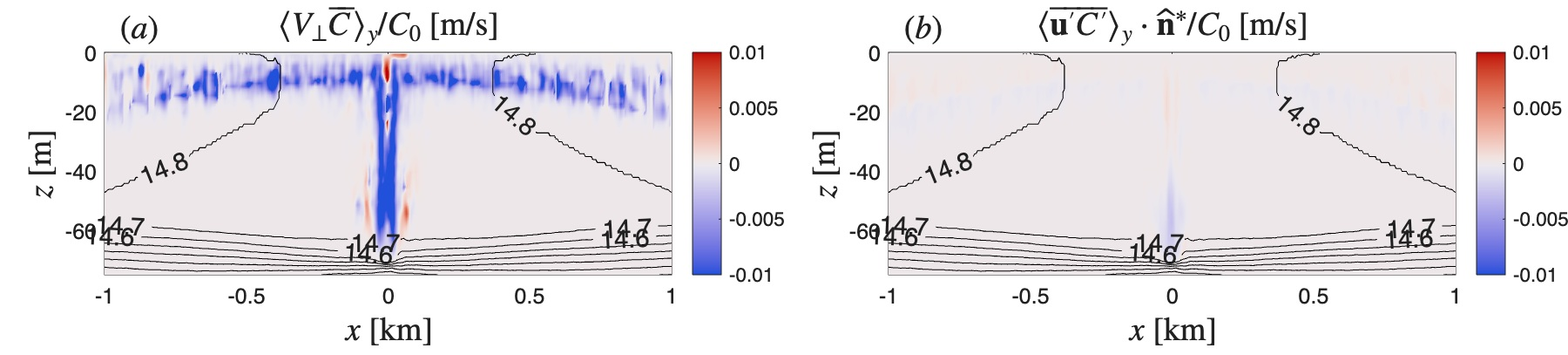}
  \caption{Side views of diapycnal tracer fluxes around peak frontal strength. 
  $(a)$: Advective diapycnal tracer flux, overlaid with temperature contours. 
  $(b)$: Turbulent tracer flux, overlaid with temperature contours.}
\label{fig:Pcon-flux}
\end{figure}

In this subsection, we analyze diapycnal tracer fluxes 
to investigate the mechanisms driving tracer transport 
into the ocean interior. The contributions from the mean 
diapycnal velocity and turbulent fluctuations are compared 
for the frontal case (figure~\ref{fig:Pcon-flux}$a,b$). 
These correspond to the advective and turbulent diapycnal 
fluxes, respectively, i.e., the second and third terms 
in equation~\eqref{eq:CV-iso}. While turbulence plays a 
crucial role in enhancing diapycnal transport, its 
influence is not primarily through the direct turbulent 
flux. Instead, the dominant pathway is the advective 
diapycnal flux, associated with the mean diapycnal 
velocity generated by turbulent density mixing. 
The SGS flux is negligible and therefore not shown. 

To quantitatively relate these diapycnal fluxes to 
the total amount of tracer transported below the mixed 
layer, we use the mass conservation equation in a 
deformable control volume (equation~\eqref{eq:CV-iso}). 
The control volume $C\V^*$ is defined as the stratified 
ocean interior, bounded above by the mixed layer base 
and extending to the domain boundaries. 
Periodic horizontal boundaries and the no-flux bottom 
boundary do not contribute to the tracer budget, 
allowing equation~\eqref{eq:CV-iso} to be rewritten as 
\begin{equation}
    \label{eq:CV-iso-replaced}
   \diff{}{t}\left(1-F_{OML}\right){ \V_0 C_0} = 
   - \int_{CS^*_{OML}} V_\perp \overline C \infd A^* 
   - \int_{CS^*_{OML}} \overline{\boldsymbol{u'}C'}\cdot\hat{n}^*\infd A^*. 
\end{equation}
Here $CS^*_{OML}=\{(x,y,z)\mid z = -h_{{OML}}(x,y,t)\}$ 
denotes the base of the ocean mixed layer. 
Integrating in time yields a decomposition of the 
fraction of tracer penetrated into the ocean interior, 
\begin{equation}
\label{eq:CV-iso-phi}
1 - F_{OML}(t) = \Phi_{adv}(t) + \Phi_{turb}(t). 
\end{equation}
The non-dimensionalized cumulative advective diapycnal 
flux is 
\begin{equation}
\label{eq:phi-adv}
\Phi_{adv}(t) = -\frac{1}{\V_0 C_0} \int_0^t \int_{CS^*_{OML}} 
V_\perp \overline{C} \infd A^* \infd t,
\end{equation}
and the cumulative turbulent flux is 
\begin{equation}
\label{eq:phi-turb-formal}
\Phi_{turb}(t) = -\frac{1}{\V_0 C_0} \int_0^t \int_{CS^*_{OML}} 
\overline{\boldsymbol{u'}C'}\cdot\hat{n}^*\infd A^* \infd t.
\end{equation}

\begin{figure}
\centering
\includegraphics[width=0.6\textwidth]{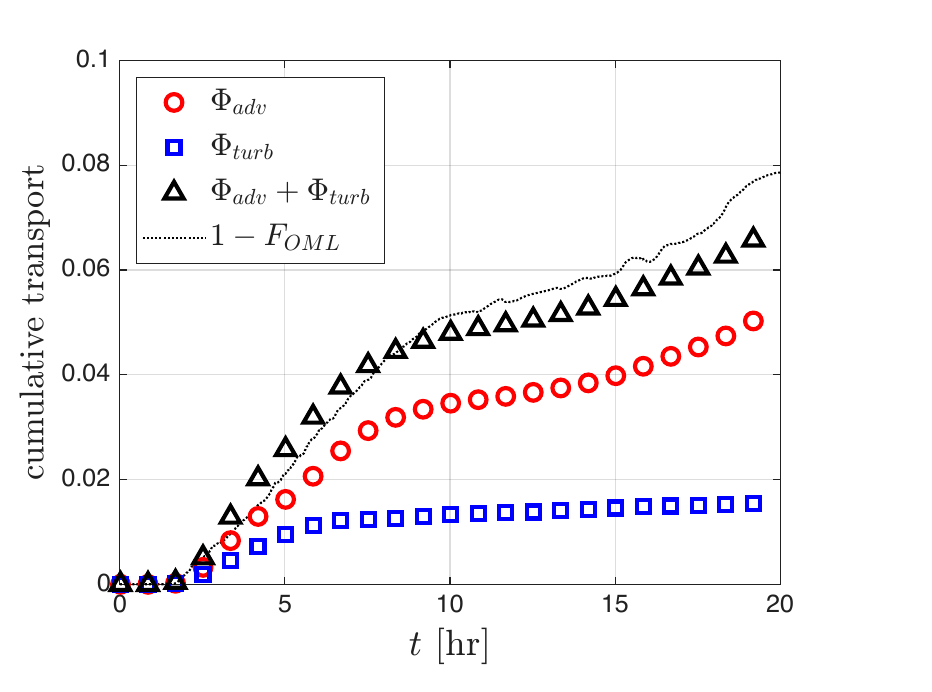}
\caption{Cumulative diapycnal fluxes as a function 
of time. See the legend for details. 
The fraction of tracer transported into the ocean 
interior ($1-F_{OML}$) is shown as a comparison.}
\label{fig:Pcon-time-fluxes}
\end{figure}

The cumulative diapycnal fluxes are evaluated in 
figure~\ref{fig:Pcon-time-fluxes}. The sum of the 
integrated flux terms ($\Phi_{adv}+\Phi_{turb}$)  
in general matches the directly calculated interior 
tracer fraction ($1-F_{OML}$), demonstrating the 
consistency of the flux decomposition. The advective 
diapycnal flux is the dominant contribution to tracer 
transport into the ocean interior, and it occurs 
as the secondary circulation velocity non-adiabatically 
crosses the isopycnals and generates $V_\perp$. 
This advective transport is most significant near peak 
frontal strength, from frontogenesis to the early stage 
of frontal decay. In contrast, the direct contribution 
from turbulent transport via small-scale eddy covariance 
is less important. 

It is worthwhile to note that the relative roles of 
advective and turbulent fluxes may depend on the 
relative distribution of tracer and density. 
When tracer gradients are not aligned with 
strong density gradients, turbulent tracer fluxes can be 
weak even under strong turbulent density mixing, allowing 
advective diapycnal flux to dominate. In contrast, 
when the initial tracer field closely follows the 
density field, turbulent tracer fluxes can make a 
more substantial contribution to diapycnal transport. 
Nevertheless, since the mean diapycnal velocity $V_\perp$ 
can reach magnitudes comparable to the turbulent velocity 
fluctuations $TKE^{1/2}$ and the secondary circulation 
velocity $\overline w$ (figures~\ref{fig:vel} and 
\ref{fig:vel-deep}), strong advective diapycnal transport 
is still expected in many submesoscale frontal zones.

\subsection{Increased mixed layer depth}
\begin{figure}
\centering
  \includegraphics[width=\textwidth]{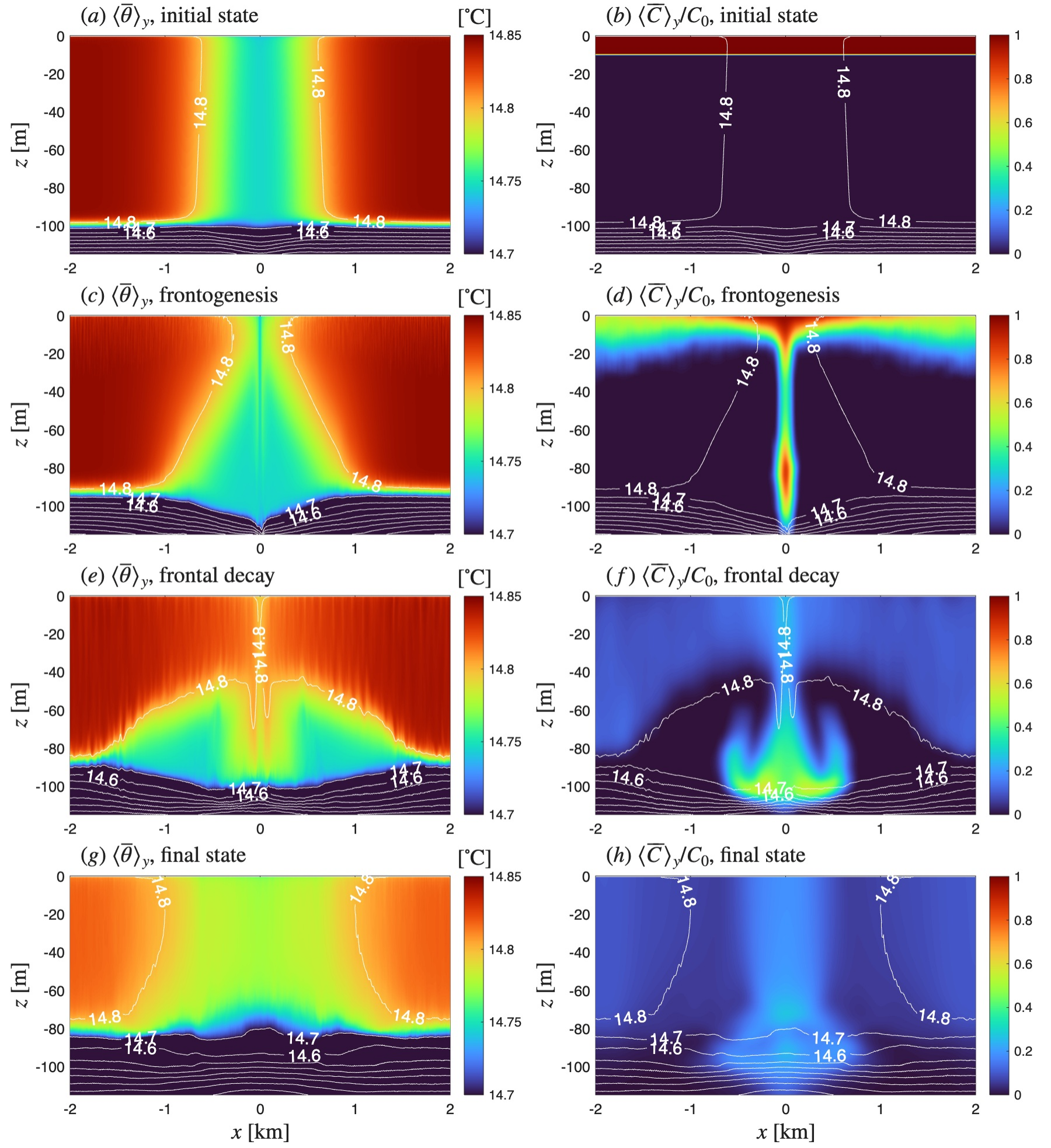}
  \caption{Side views of the along-front averaged temperature 
  (left column) and tracer concentration (right column) fields 
  in the frontal case with a deep mixed layer (case SD). 
  $(a)$, $(c)$, $(e)$, and $(g)$: Temperature fields at four 
  representative times: the initial state, the frontogenesis phase, 
  the early stage of frontal decay, and the final state, respectively. 
  $(b)$, $(d)$, $(f)$, and $(h)$: Tracer distributions at the 
  same four times, with tracer initially released across the 
  entire horizontal domain near the sea surface. 
  Tracer concentration is normalized by the initial concentration $C_0$.}
\label{fig:temp-deep}
\end{figure}

Submesoscale currents in the real ocean span a wide range of 
strengths, geometries, and background conditions. The analysis 
presented above focuses on an idealized filament configuration 
designed to isolate the key mechanisms governing diapycnal 
tracer transport. In this section, we further examine the 
robustness of the identified transport pathways and evaluate 
the extent to which the proposed mechanism applies across 
different submesoscale regimes. 

\begin{figure}
\centering
  \includegraphics[width=\textwidth]{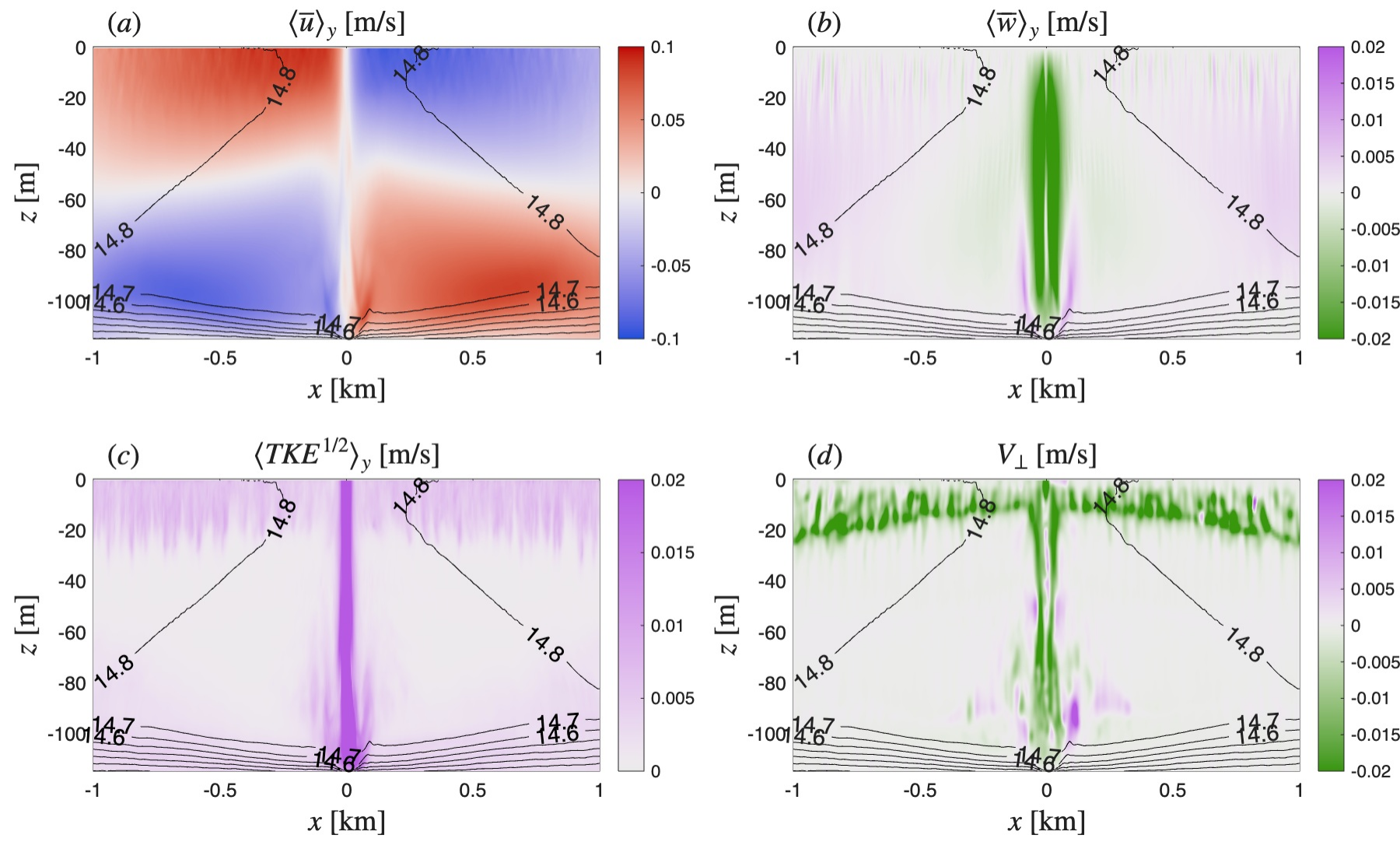}
  \caption{Side views of mean and turbulent velocity fields 
  in the frontal case with a deep mixed layer 
  (case SD). Results are averaged in the 
  along-front direction and overlaid with temperature contours.
  $(a)$: Mean cross-front velocity. 
  $(b)$: Mean vertical velocity. 
  $(c)$: Square root of the turbulent kinetic energy, 
  representing a turbulent velocity scale. 
  $(d)$: Mean diapycnal velocity. 
  Negative values indicate flow toward lower-temperature regions.}
\label{fig:vel-deep}
\end{figure}

We first present results for the deep frontal case with 
the same tracer release setup as in the baseline shallow 
mixed layer frontal case, in which tracer is uniformly 
released near the sea surface across the entire horizontal 
domain (figure~\ref{fig:temp-deep}, case deep frontal SD). 
Similar rapid frontogenesis and turbulence intensification 
are observed, driving strong diapycnal tracer transport 
into the stratified interior. The frontal strength reaches 
its peak slightly earlier than in the shallow mixed layer 
case, as the stronger secondary circulation induces 
enhanced horizontal convergence and a larger frontogenetic 
tendency at the filament center. 
The transport mechanism involving mean diapycnal 
velocity, associated with secondary circulation and 
turbulent density mixing, remains active in the deep mixed 
layer frontal case, and the resulting advective flux 
continues to play a dominant role in transporting tracer 
into the ocean interior (figure~\ref{fig:vel-deep}). 

A second configuration is examined in which tracer is 
initially released throughout most of the mixed layer 
($z=-80$~m to $z=0$~m, case deep frontal MD), 
rather than being confined near the surface 
as in the previous cases ($z=-10$~m to $z=0$~m). 
Results for the mixed-layer release are not shown, but 
the tracer transport mechanism and the enhanced diapycnal 
fluxes remain similar.

\begin{figure}
\centering
  \includegraphics[width=0.6\textwidth]{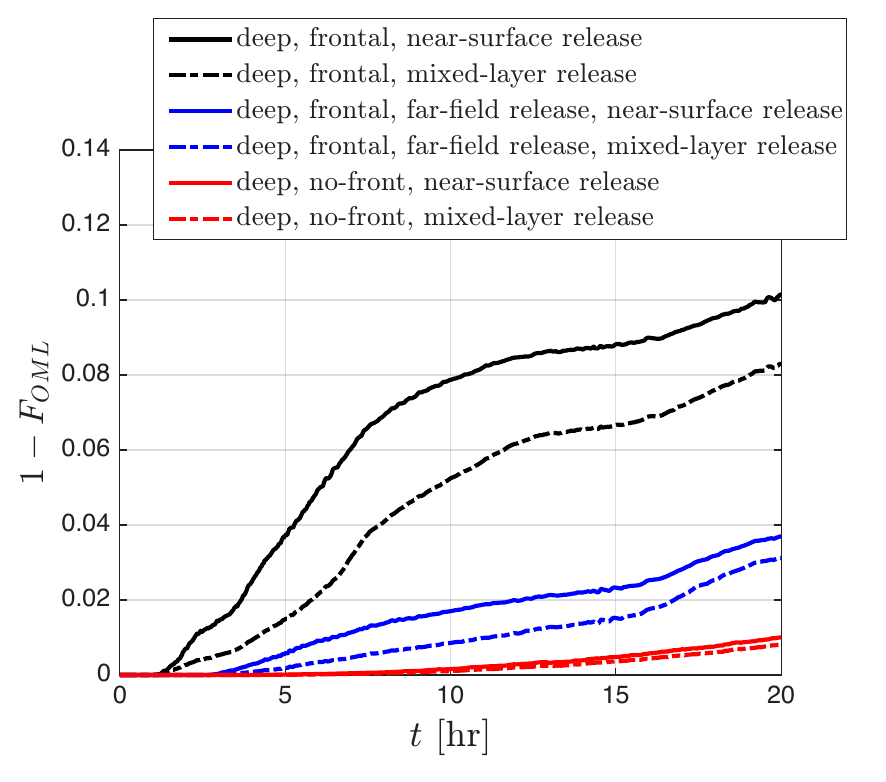}
  \caption{Time series of the fraction of tracer that has penetrated 
  below the ocean mixed layer. Lines show the frontal and no-front cases 
  with a deep mixed layer, for different tracer release configurations; 
  see the legend for details.}
\label{fig:Pcon-time-deep}
\end{figure}

While strong diapycnal transport occurs in the frontal 
case with a deep mixed layer, tracer flux into the stratified 
interior remains weak in the corresponding no-front case, 
where only convective turbulence develops in the absence of 
frontogenesis. Tracer transport is also quantified for the 
deep mixed layer simulations using \eqref{eq:tracer-fraction}. 
A substantially larger fraction of tracer penetrates into the 
ocean interior in the frontal case than in the no-front case 
with a deep mixed layer (figure~\ref{fig:Pcon-time-deep}), 
demonstrating that the enhancement of diapycnal transport by 
frontogenesis persists in the deeper mixed layer. 

\begin{figure}
\centering
  \includegraphics[width=\textwidth]{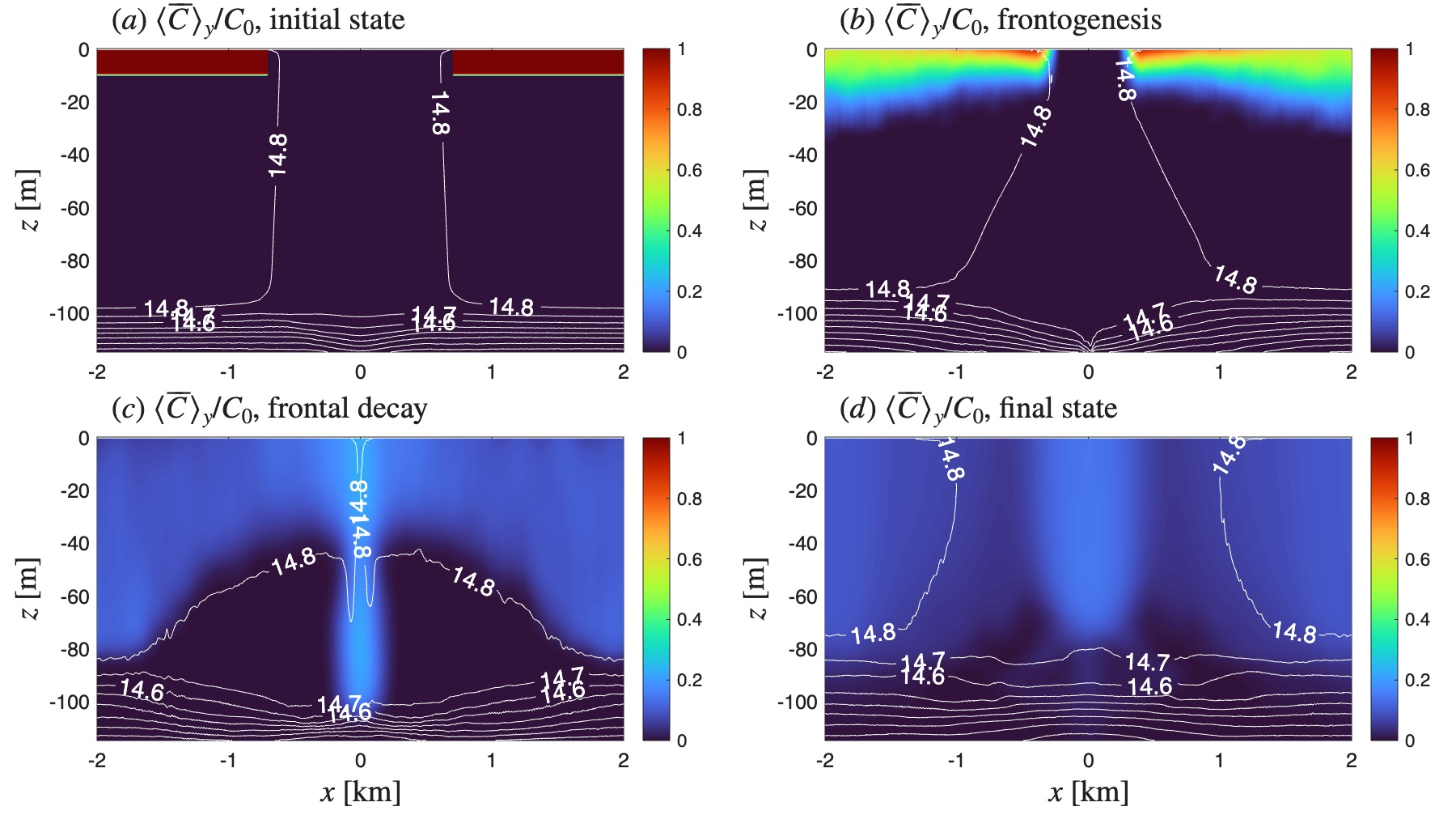}
  \caption{Side views of the along-front averaged tracer 
  concentration field in the frontal case with a deep mixed 
  layer, with tracer initially released in the far field, 
  i.e., away from the filament center (case SF). 
  $(a)$, $(b)$, $(c)$, and $(d)$: The initial state, the 
  frontogenesis phase, the early stage 
  of frontal decay, and the final state, respectively.}
\label{fig:temp-deep-side}
\end{figure}

\begin{figure}
\centering
  \includegraphics[width=\textwidth]{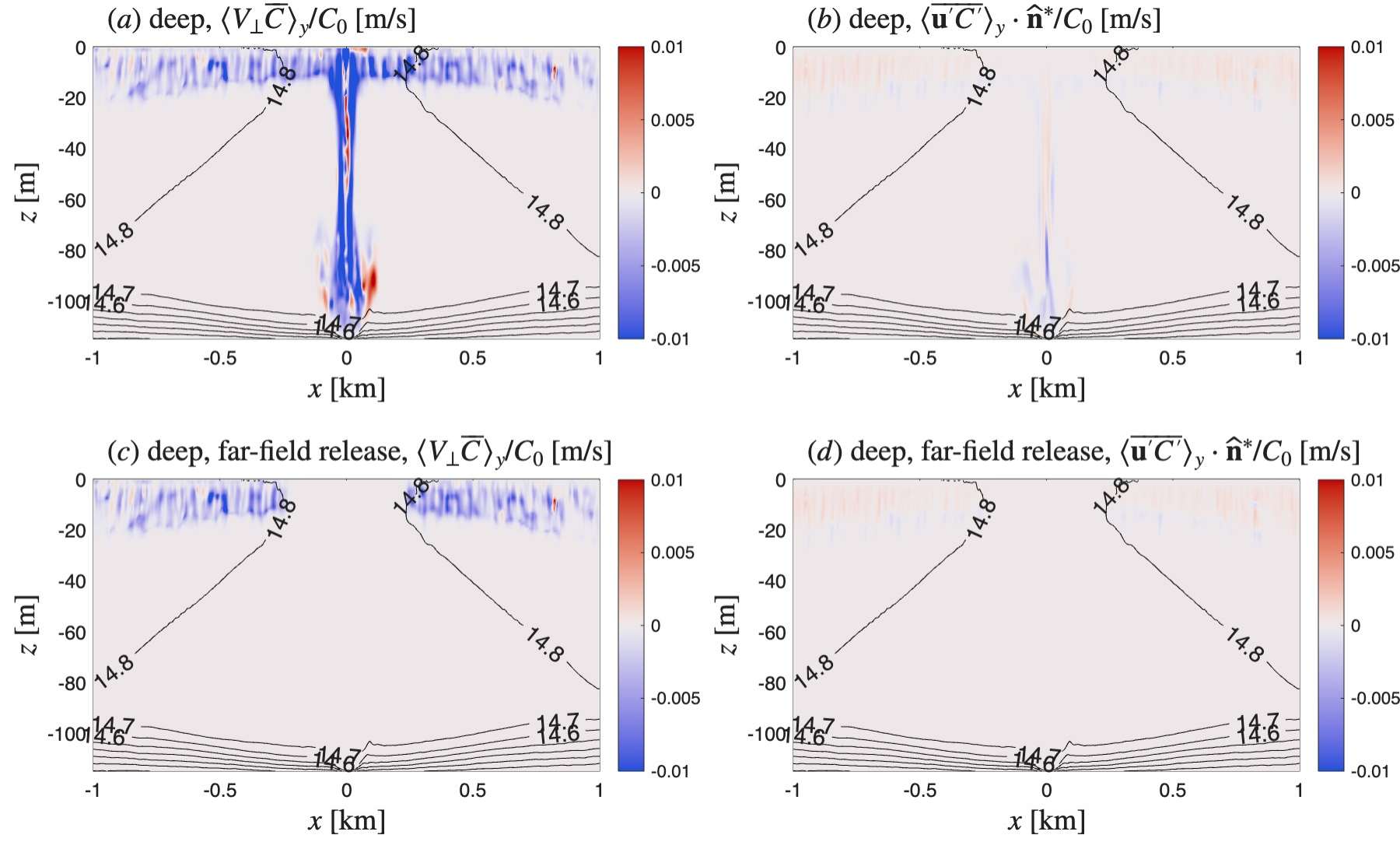}
  \caption{Side views of diapycnal tracer fluxes 
  in cases with a deep mixed layer. 
  $(a)$ and $(b)$: The deep frontal case with 
  domain-wide tracer release (case SD).
  $(c)$ and $(d)$: The deep frontal case with 
  far-field tracer release (case SF).
  $(a)$ and $(c)$: Advective diapycnal tracer flux. 
  $(b)$ and $(d)$: Turbulent tracer flux.}
\label{fig:Pcon-flux-deep}
\end{figure}

We also consider cases in which tracer is released only 
in the far field, away from the filament center 
(cases deep frontal SF and MF). 
In this setup, additional time is required for the tracer 
to advect toward the filament center, where strong 
diapycnal fluxes occur (figure~\ref{fig:temp-deep-side}). 
As a result, the released tracer does not fully experience 
the early stage of frontogenesis, during which turbulence 
is intensified. Consequently, the diapycnal fluxes are 
weaker in the far-field release case than in the domain-wide 
release case (Figure~\ref{fig:Pcon-flux-deep}). 
Overall, a smaller fraction of tracer is transported below 
the mixed layer in the far-field release case, 
yet diapycnal transport is still stronger than in the 
no-front case (figure~\ref{fig:Pcon-time-deep}). 

These results indicate that the enhancement of diapycnal 
transport is transient, since frontogenesis and turbulence 
intensification both develop over a relatively short time 
period. Nevertheless, enhanced diapycnal transport still 
occurs once the tracer encounters regions of intense frontal 
turbulence. It is important to note that, although the 
enhanced transport occurs rapidly during an individual 
frontogenesis event, the conditions that give rise to it 
occur repeatedly under a wide range of ocean conditions. 
Therefore, the associated tracer transport is expected to 
play a recurring role in enhancing vertical material exchange 
in the upper ocean. 

In addition to mixed layer depth, other factors such as 
the initial frontal strength and surface boundary forcing 
also affect frontal dynamics and turbulence generation 
\citep{sullivan2019,sullivan2024}. 
The influences of these factors on diapycnal tracer 
transport may therefore be inferred, since diapycnal fluxes 
are closely linked to frontal turbulence and the 
downwelling of secondary circulation. 
In our post-frontal case, the decayed front has a strength 
comparable to that of the weakest fronts examined in 
\citet{sullivan2019} and \citet{sullivan2024}; thus, 
the frontal and post-frontal cases could effectively 
bracket their typical range of frontal strengths. 
In addition, while this study focuses on convective 
surface cooling as the only boundary forcing, 
more complex surface forcing such as wind stress or 
surface waves may disrupt the symmetry of filament shape 
as well as the associated tracer transport.

\section{Discussion}
\label{sec:discussion}
\subsection{Tracer release in the ocean interior}

\begin{figure}
\centering
  \includegraphics[width=\textwidth]{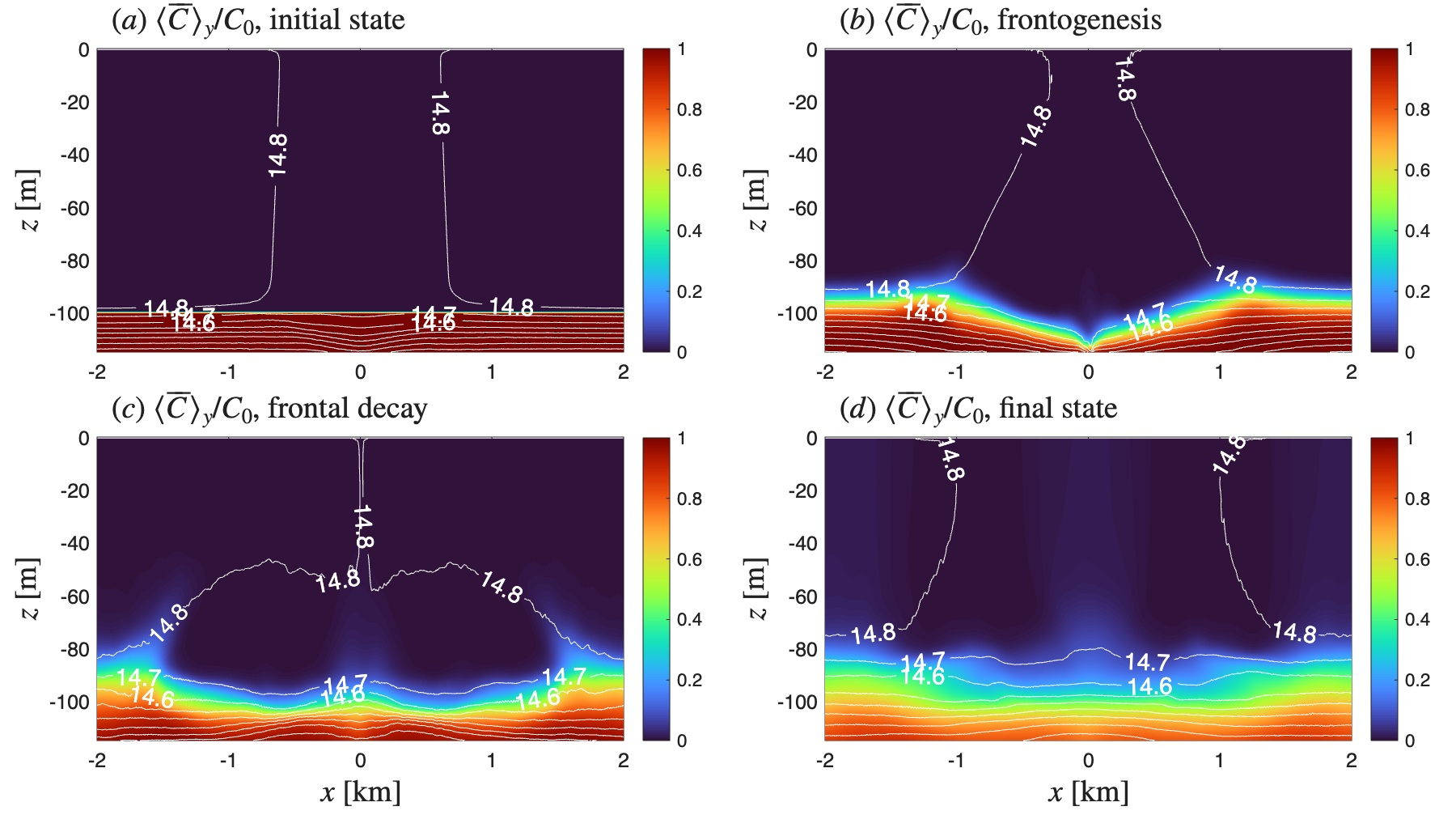}
  \caption{Side views of the along-front averaged 
  tracer concentration field in the frontal case with 
  a deep mixed layer, with tracer initially released 
  in the ocean interior. 
  $(a)$, $(b)$, $(c)$, and $(d)$: The initial state, the 
  frontogenesis phase, the early stage 
  of frontal decay, and the final state, respectively. }
\label{fig:temp-deep-bottom-up}
\end{figure}

This study focuses on tracer transport from the surface 
mixed layer into the stratified interior. 
The transport pathway may differ when tracer is instead 
released within the stratified interior, which merits 
further investigation. 
To provide an initial assessment 
of this, we conduct a simulation for the deep mixed layer 
frontal case in which tracer is released within the 
stratified interior (figure~\ref{fig:temp-deep-bottom-up}, 
case deep frontal I). 
Turbulent flux (eddy covariance) contributes to upward 
tracer transport in the filament center. 
In the meantime, upward transport also occurs on the filament 
flanks, i.e., in lateral regions away from the filament 
center where secondary circulation drives upwelling. 
It is also worth noting that turbulent flux can be 
vertically asymmetric, i.e., transport from the surface 
to the interior and from the interior to the surface can 
differ in magnitude, depending on turbulence 
characteristics \citep{chor2020,chor2021}.

\subsection{Single-sided fronts}

\begin{figure}
\centering
  \includegraphics[width=\textwidth]{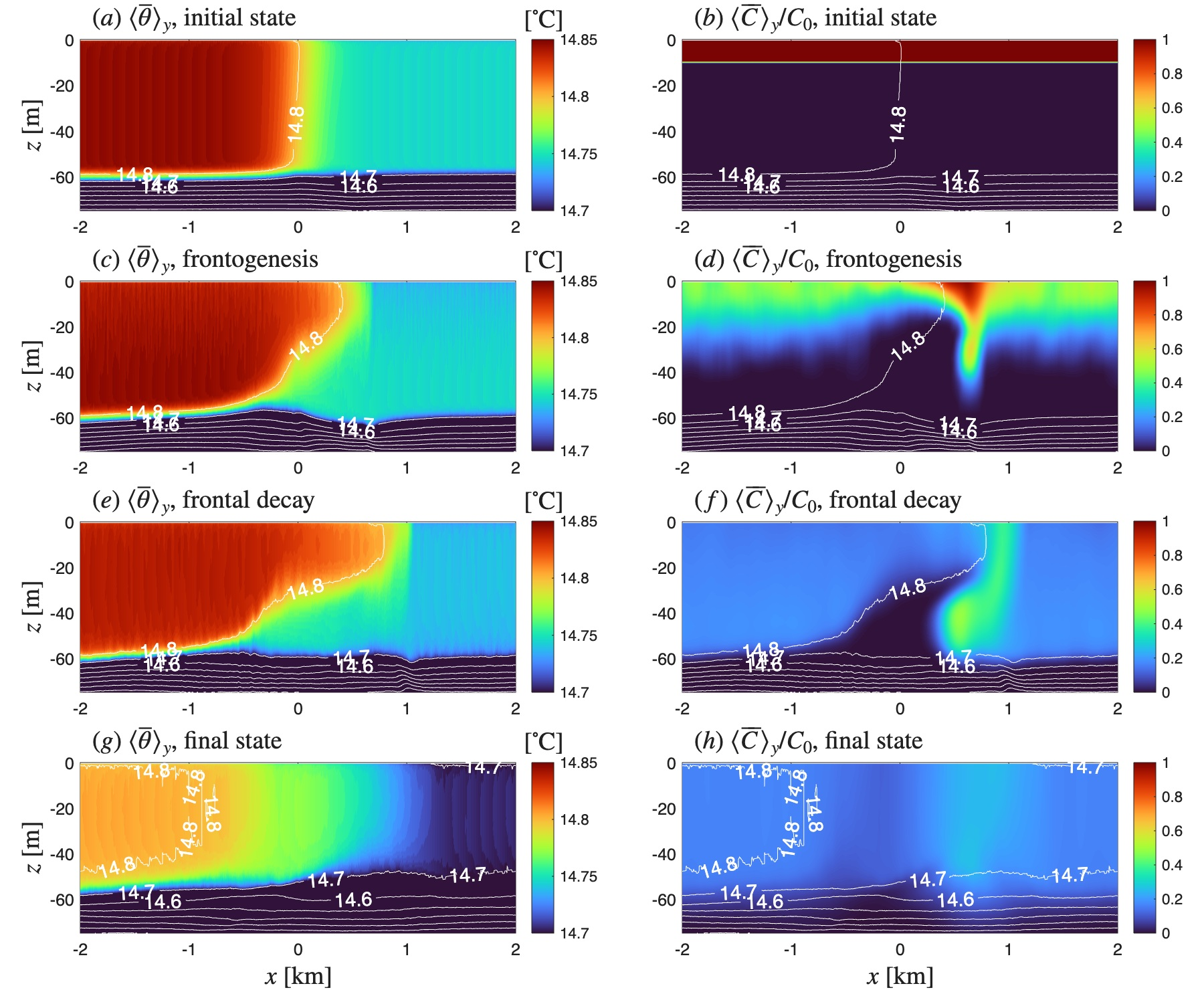}
  \caption{Side views of the along-front averaged 
  temperature and tracer concentration fields 
  in the simulation with a single-sided front. 
  $(a)$, $(c)$, $(e)$, and $(g)$: Temperature fields 
  at four representative times: the initial state 
  ($t=0$), the frontogenesis phase ($t=1.5$~hr), 
  the early stage of frontal decay ($t=5$~hr), and 
  the final state ($t=20$~hr), respectively. 
  $(b)$, $(d)$, $(f)$, and $(h)$: Tracer distributions 
  at the same four times, with tracer initially released 
  across the entire horizontal domain near the sea surface.}
\label{fig:temp-single}
\end{figure}

\begin{figure}
\centering
  \includegraphics[width=0.6\textwidth]{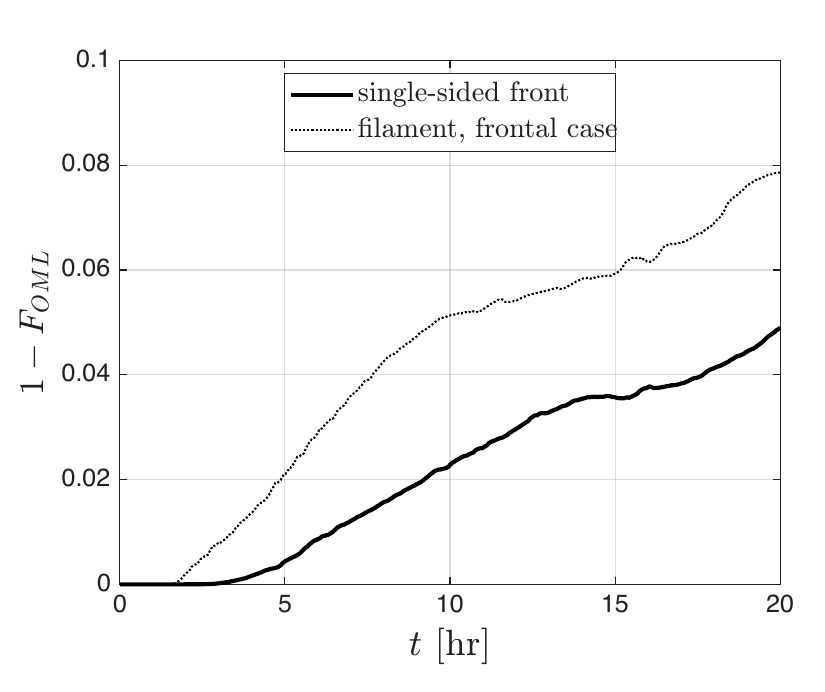}
  \caption{The fraction of tracer that has penetrated 
  below the ocean mixed layer in the single-sided front simulation.}
\label{fig:Pcon-time-single}
\end{figure}

We have focused on a dense filament in the previous 
sections, which can be viewed as a two-sided submesoscale 
front. As an additional comparison, we also investigate 
a single-sided front to explore tracer transport across 
a broader range of submesoscale frontal configurations. 
The single-sided front simulation uses the same domain 
size, grid resolution, surface forcing, initial frontal 
strength ($M^2=50$), and mixed layer depth ($h_{0,OML}=55$~m) as 
the baseline filament case analyzed above. 
The only difference is that the initial field consists 
of a single density step, rather than a symmetric 
filament structure characterized by density gradients on 
both sides of a density maximum. 
A background density gradient method is applied in 
the single-sided front simulation \citep{bo2025front} 
to ensure that the periodic boundary condition is 
satisfied in the cross-front direction ($x$-direction). 

The single-sided front undergoes similar frontogenesis 
and turbulence intensification, driving diapycnal tracer 
transport into the ocean interior (figure~\ref{fig:temp-single}). 
This further demonstrates the generality of enhanced 
tracer fluxes associated with submesoscale frontogenesis. 
However, single-sided fronts typically have weaker frontal 
strength than filaments, even with identical initial conditions 
and surface forcing \citep{mcwilliams2009,bo2025front}. 
Accordingly, the overall fraction of tracer penetrating into 
the ocean interior is smaller (figure~\ref{fig:Pcon-time-single}). 
Moreover, while the filament remains stationary because of its 
symmetric structure, the single-sided front propagates toward 
the denser side. As a result, the region of strongest diapycnal 
transport moves with the frontal zone, causing the subducted 
tracer to become displaced from the surface front 
(figure~\ref{fig:temp-single}$f$).

\section{Conclusion}
\label{sec:conclusion}
This study uses LES to investigate tracer transport 
associated with submesoscale fronts. Intense turbulence 
is generated as a result of frontogenesis, which far exceeds 
background turbulence levels in the surface mixed layer. 
Diapycnal tracer transport occurs in the frontal zone, 
enhancing material exchange between the surface and the 
ocean interior. While turbulence is widely recognized as a key 
driver of diapycnal transport, our findings suggest that the 
dominant pathway here is not the direct turbulent tracer flux. 
Instead, submesoscale frontogenesis leads to a strong 
advective diapycnal flux, driven by a mean diapycnal velocity 
induced by turbulent mixing of density. 
Given the prevalence of submesoscale fronts in the global 
ocean, the identified mechanism may play a critical role 
in shaping vertical tracer distributions. It thus merits 
increased attention in climate modeling and biogeochemical 
cycling studies and warrants further exploration under 
varying ocean conditions. 

Several previous LES studies of submesoscale fronts 
have examined turbulence and tracer transport 
\citep[e.g.,][]{hamlington2014,smith2016}, but their initial 
conditions often lacked coexisting frontal structures and 
turbulence. As a result, these simulations typically 
emphasize baroclinic or symmetric instabilities, and 
frontogenesis may be delayed or even inhibited when 
turbulence spins up from rest \citep{sullivan2018,bo2025front}. 
Our LES study directly resolves tracer transport during 
active frontogenesis and turbulent mixing, highlighting 
the need for improved representation of these processes 
in traditional ocean models. 

As shown in the reference simulation where 
tracer is released after frontal decay, a substantial portion 
of diapycnal tracer transport is missing. This likely reflects 
the level of frontal dynamics that traditional ocean models are 
able to resolve, where weak frontal structures and secondary 
circulation exist, but the most active periods of frontogenesis, 
turbulence generation, and frontal downwelling are not captured. 
Accurately representing frontal turbulence and tracer 
transport in ocean models faces two major challenges: 
first, high grid resolution is required to characterize 
submesoscale structures and localized tracer fluxes, as the 
frontal zone can narrow to less than 100~m; second, improved 
turbulence closures are needed that simultaneously account 
for momentum, density, and passive tracers, since diapycnal 
transport depends on their coupled effects.

\backsection[Acknowledgements]{
The authors acknowledge computational support from 
Derecho (doi.org/10.5065/qx9a-pg09) 
provided by NCAR’s Computational and Information Systems 
Laboratory, sponsored by the National Science Foundation, 
and support from High-performance Computing Platform 
of Peking University. }

\backsection[Funding]{The authors acknowledge financial 
support from the National Science and Technology Major 
Project of China (Grant No. 2025ZD1403007) and the 
U.S. Department of Energy ARPA-E MARINER 
program (Grant No. DE-AR0000920).}

\backsection[Data availability statement]{The data that support the 
findings of this study are openly available on Zenodo at 
https://doi.org/10.5281/zenodo.20339306.}

\backsection[Declaration of interests]{The authors report no conflict of interest.}

\bibliographystyle{jfm}
\bibliography{references}

\end{document}